\documentclass[sn-mathphys]{sn-jnl}

\jyear{2021}%

\theoremstyle{thmstyleone}%
\theoremstyle{thmstyletwo}%
\newtheorem{remark}{Remark}%

\theoremstyle{thmstylethree}%

\newcommand{\meta}{{MetaNOR}}
\usepackage{bibentry}
\usepackage{bm}

\newcommand{\vb}{{\mathbf{b}}}

\newcommand{\real}{\mathbb{R}}

\newcommand{\mcD}{\mathcal{D}}

\newcommand{\mcF}{\mathcal{F}}

\newcommand{\mcJ}{\mathcal{J}}
\newcommand{\mcL}{\mathcal{L}}

\newcommand{\mcR}{\mathcal{R}}

\newcommand{\mcU}{\mathcal{U}}

\usepackage{pifont}

\def\omg{{\Omega}}

\def \Bb{\mathbf{B}}

\def \Cb{\mathbf{C}}

\def \vb{\mathbf{v}}

\def \Bb{\mathbf{B}}

\newcommand{\vertii}[1]{{\left\vert\left\vert #1
    \right\vert\right\vert}}

\newcommand{\verti}[1]{{\left\vert #1
    \right\vert}} 
    
\newtheorem{counter}{Counter}[section]
\newtheorem{lem}[counter]{Lemma}

\newtheorem{thm}[counter]{Theorem}

\newtheorem{asp}[counter]{Assumption}

\begin{document}

\title[MetaNOR: A Meta-Learnt Nonlocal Operator Regression Approach]{MetaNOR: A Meta-Learnt Nonlocal Operator Regression Approach for Metamaterial Modeling}


\author[1]{\fnm{Lu} \sur{Zhang}}\email{luz319@lehigh.edu}

\author[1]{\fnm{Huaiqian} \sur{You}}\email{huy316@lehigh.edu}

\author*[1]{\fnm{Yue} \sur{Yu}}\email{yuy214@lehigh.edu}

\affil*[1]{\orgdiv{Department of Mathematics}, \orgname{Lehigh University},\\\orgaddress{\street{27 Memorial Dr W}, \city{Bethlehem}, \postcode{18015}, \state{PA}, \country{USA}}}


\abstract{
We propose MetaNOR, a meta-learnt approach for transfer-learning operators based on the nonlocal operator regression. The overall goal is to efficiently provide surrogate models for new and unknown material-learning tasks with different microstructures. The algorithm consists of two phases: (1) learning a common nonlocal kernel representation from existing tasks; (2) transferring the learned knowledge and rapidly learning surrogate operators for unseen tasks with a different material, where only a few test samples are required. We apply MetaNOR to model the wave propagation within 1D metamaterials, showing substantial improvements on the sampling efficiency for new materials.}




\maketitle


\section{Introduction}

\textit{Metamaterial} is a group of artificial heterogeneous materials exhibiting unusual yet desired frequency dispersive properties from its composite microstructure. 
In the past decade, there have been increasing research interests in metamaterials in scientific and engineering communities, because of its applications in acoustic attenuation, noise control, and invisibility cloaking \cite{li2019design}. 

\begin{figure}[t]
\centering
\includegraphics[width=1.0\columnwidth]{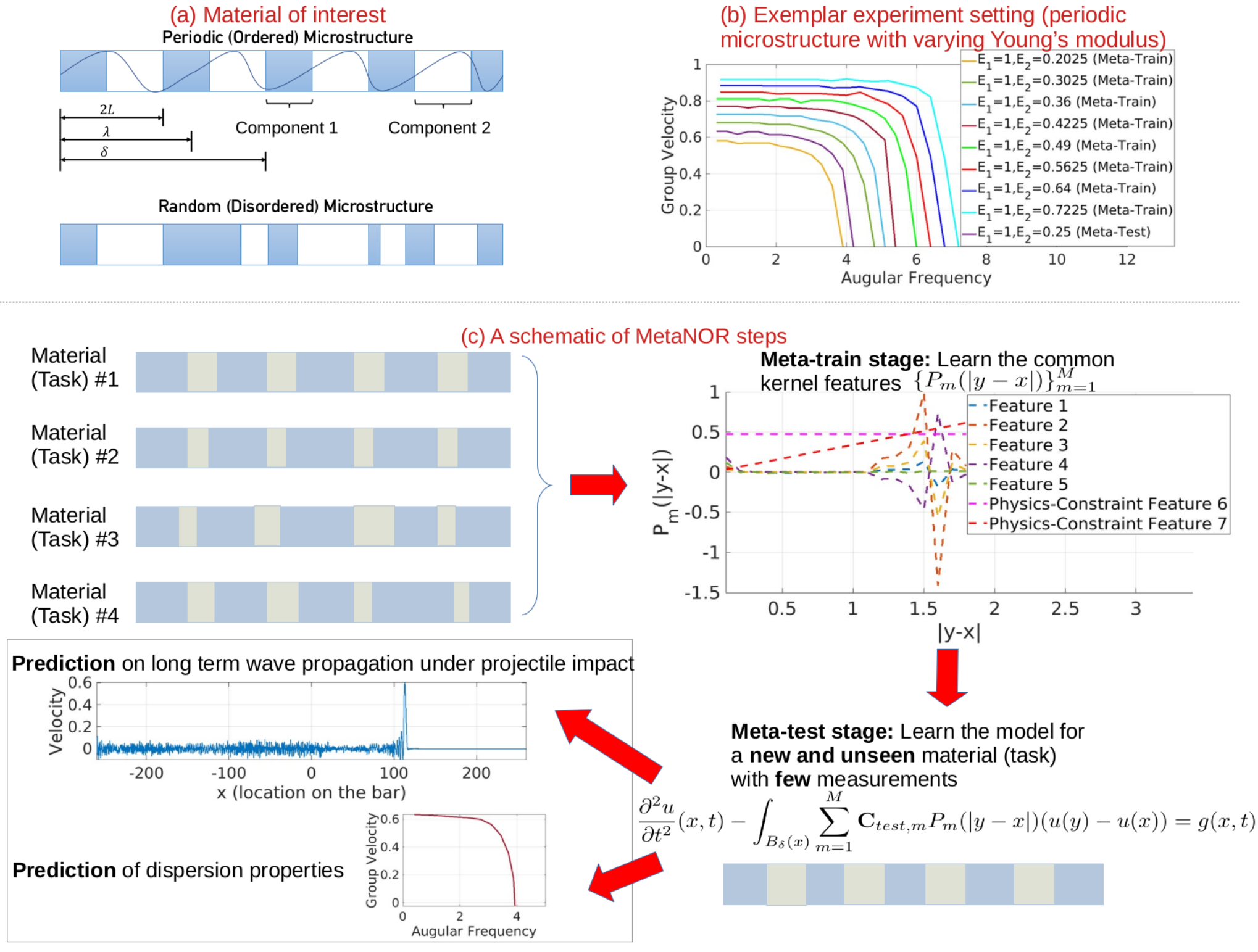}
\caption{Problem of interests and a schematic of the proposed algorithm. (a) One-dimensional metamaterial composed by dissimilar components 1 and 2. Components 1 and 2 have the densities $\rho_1$ and $\rho_2$ and Young's modulus $E_1$ and $E_2$. The horizon $\delta$, and the wave length $\lambda$, are reported for comparison. (b) A demonstration of our experiment setting. (c) A schematic of the main steps in MetaNOR.}
\label{fig:bar}
\end{figure}

Metamaterials has been studied theoretically, numerically and experimentally \cite{yao2008experimental,huang2009wave,manimala2014dynamic}, to optimize its microstructure for dispersive properties and performances in different environments. However, their design and modeling is  often computationally prohibitive, since  they contain more than thousands of interfaces in its microstructures (example interfaces are shown in Figure~\ref{fig:bar}). For example,  modeling dispersive properties via the stress wave propagation in metamaterials would require accurate bottom-up characterization of material interfaces and simulations from each individual layer (the microscopic scale), which are often of orders smaller than the problem domain length scale (the macroscopic scale). Therefore, even with sophisticated optimization techniques, designing such a material would still require running  multiple full wave simulations for each candidate microstructure and consumes significant computation time. 


To accelerate stress wave simulations, efficient surrogate models for metamaterials such as homogenized models are often employed  \cite{du2019multiscale}. They
are posed as a single equation of the displacement field and can be readily used in simulations at the macroscopic scale.
Conventional homogenized surrogate models are proposed as continuum partial differential equation (PDE) models \cite{barker1971model},  from mathematical derivations or learned methods with data. 
However, typical homogeneous elastic model within the PDE theory does not account for wave dispersion, and therefore is often valid only in the limiting case of a very small length scale in the oscillatory behavior of the original material parameters. In some applications, the classical homogenization theory may ``washed out'' the length scale in the original problem, causing information to be lost. 

Recently, nonlocal surrogate models have also received lots of attentions \cite{You2021,silling2021propagation,deshmukh2021}, 
where integral operators are employed which embeds all time and length scales in their definitions. Comparing with continuum PDE counterparts, the nonlocal surrogate models provide a more natural affinity with the dispersive waves in a microstructure, and successfully reproduce many features of the decay and spreading of stress waves \cite{deshmukh2021}. Moreover, in nonlocal homogenization models the choice of kernel functions contains information about the small-scale response of the system. Hence, once equipped with the power of machine learning to identify the optimal form for the kernel function, effective nonlocal surrogate models can be obtained from high-fidelity simulations and/or experimental measurements, so as to reproduce the material response with the greatest fidelity. To this end, the nonlocal operator regression (NOR) approach \cite{You2020Regression,You2021,you2021md} is proposed to obtain large-scale nonlocal descriptions and capture small scale material behavior that would remain hidden in classical approaches to homogenization. 
 NOR and the general homogenized surrogate models have greatly accelerate the simulation of heterogeneous materials at macroscale, but the choice of homogenized models is often selected case by case, which makes rigorous model calibration for multiple microstructure challenging and time consuming. Moreover, in metamaterials and the general heterogeneous material modeling problems, the data acquisition is often very challenging and expensive, which makes it critical for any method to learn the material model with a limited number of measurements.

Motivated by further accelerating the design process for metamaterials, in this paper we leverage NOR and the general heterogeneous material homogenization procedure, to answer the following question: \textit{Given  knowledge on a number of materials with different microstructures, how can one efficiently learn the best surrogate model for a new and possibly unknown microstructure, with only a small set of training data (such as several pairs/measurements of displacement and loading fields from experiments)?}

To answer this question, we  develop sample-efficient data-driven homogenized models for new metamaterial with unseen microstructure, which (1) allow for accurate simulations of wave propagation at much larger scales than the microstructure; (2) provide constitutive laws that can be readily applied in simulation problems posed on various environments with general geometries and complex time-dependent loadings; and (3) utilize the knowledge from previously studied materials to rapidly adapt to new microstructures. 
Specifically, inspired by a recent provable meta-learning approach of linear representations \cite{tripuraneni2021provable}, we propose meta nonlocal operator regression (\meta), a sample-efficient learning algorithm for metamaterials where multiple tasks (microstructures) share a common set of low-dimensional features in their kernel space, for accurate and efficient adaptation to unseen tasks.  
The algorithm has three components. First, in the \textit{meta-train} stage, we learn a common set of features in the nonlocal kernel space from multiple related tasks (i.e., related microstructures), by minimizing the corresponding empirical risk induced by  nonlocal surrogate models. 
Second, in the \textit{meta-test} stage, with estimated kernel feature representation shared across tasks, we transfer this knowledge to learn the model for a new and unseen microstructure, where only a few samples/measurements are required. Third, when partial physical knowledge is available, such as the effective wave speed for infinitely long wavelengths and/or the dispersion properties of the material for very long waves, we incorporate these physics-based constraints into the proposed meta-learning algorithm to further improve the learning accuracy and sample efficiency.

\noindent\textbf{Main Contributions}: we summarize the main contributions of this paper as follows:
\begin{enumerate}
    \item We design a novel meta-learning technique for learning nonlocal operators, by learning a common set of low-dimensional features on multiple known tasks in their kernel space and then transferring this knowledge to new and unseen tasks.
     \item Our method is the first application of meta-learning approach on homogenized model for heterogeneous materials, which 
     efficiently provides an associated model surrogates that are effective on applications with various loading and time scales.
    \item We provide rigorous error analysis for the proposed algorithm, showing that the estimator converges as the data resolution refines and the number of sample increases. We verify its efficacy on a synthetic dataset.
    \item We illustrate the proposed method on one-dimensional  metamaterials, to confirm the applicability of our technique and the improved sample efficiency over  baseline nonlocal operator regression estimators.
\end{enumerate}

\section{Nonlocal Operator Regression (NOR)}

We first 
review related concepts of the general nonlocal operator regression (NOR) approach, and 
then demonstrate the equivalence of NOR and a linear regression model in the kernel space, which provides the foundation for us to formulate {\meta} as a linear kernel feature learning problem with theoretical guarantees in the next section.

Through out this paper, we use $\mathbf{A}^*$ to denote its optima for each vector or matrix or operator $\mathbf{A}$ of interests, $\mathbf{A}^+$ to denote its ground-truth, and $\mathbf{A}^\circ$ to denote an approximation solution to the optima. For any vector $\vb=[v_1,\cdots,v_q]\in\real^q$, we use $\vertii{v}_{l^2}:=\sqrt{\sum_{i=1}^qv_i^2}$ to denote its $l^2$ norm, and similarly $\vertii{v}_{l^1}:={\sum_{i=1}^q\verti{v_i}}$ denotes the $l^1$ norm. For any matrix $\mathbf{A}=[A_{ij}]\in\real^{p\times q}$, we use $\vertii{\mathbf{A}}_F:=\sqrt{\sum_{i=1}^p\sum_{j=1}^q A_{ij}^2}$ to denote its Frobenius norm, and use $\vertii{\mathbf{A}}$ to denote its spectral norm. $\gtrsim$ and $\lesssim$ denote greater than and less than, up to a universal constant, respectively. We use $O$, $\Omega$ and $\Theta$ as in the standard notations, and $\tilde{O}$ as an expression that hides polylogarithmic factor in problem parameters. $\mathbf{I}_p$ denotes the $p\times p$ identity matrix. 

\subsection{Nonlocal Models and Kernel Learning}

Nonlocal Models \cite{beran1970mean}  describe the state of a system, where any point depends on the state in a neighborhood of points. 
In heterogeneous materials, it is shown that nonlocality naturally appears in the homogenized model derived from  micromechanical models \cite{silling2014origin}, which makes nonlocal operators good candidates for obtaining homogenized models for heterogeneous materials \cite{du2020multiscale}. 

Formally, on some bounded domain $\omg\in\real^p$, we model the high-fidelity or ground-truth material response as a mapping between two function spaces, 
$\mcL^+:\mcU\rightarrow\mcF$, where $\mcU=\{u(x),x\in\omg\}$ (can be seen as the space of displacement fields) and $\mcF=\{f(x),x\in\omg\}$ (can be seen as the space of loading fields) are Banach spaces. The goal of nonlocal operator learning is to then find a surrogate operator $\mcL_{\theta}:\mcF\rightarrow \mcU$ with parameter $\theta$, such that $\mcL_{\theta}\approx \mcL^+$. In particular, we assume that the action of the ground-truth material model may be approximated by an integral operator of the form:
\begin{equation}\label{eqn:intoperator}
    \mcL_{\theta}[u](x)=\int_{\omg} \gamma_{\theta}(x,y)(u(y)-u(x))dy,
\end{equation}
where the kernel $\gamma$ takes inputs spatial locations $x$ and $y$. Inspired by the application of nonlocal diffusion operator in metamaterial problems \cite{silling2021propagation,You2021}, we choose to take $\gamma_{\theta}(x,y):=\gamma_\theta(\verti{y-x})$ as a radial and sign-changing kernel function, which is compactly supported on the ball of radius $\delta$ centered at $x$, denoted as $B_\delta(x)$.  Moreover, we represent $\gamma_{\theta}$ as a linear combination of basis polynomials:
\begin{equation}\label{eqn:linear}
\gamma_{\theta}\left(\verti{y-x}\right)=\sum_{m=1}^M C_m P_{m}(\verti{y-x}),    
\end{equation}
with properly chosen basis polynomials $\{P_{m}(\verti{y-x})\}$. 

Suppose we are given observations of $K$ pairs of functions $\{u_k(x),f_k(x)\}_{k=1}^K$ where $u_k(x)$ are samples in $\mcU$ and $\mcL^+[u_k]=f_k$, potentially with noise.  Then, learning the nonlocal kernel $\gamma_\theta$ can be framed as an optimization problem. 
Here we consider the squared loss in the $L^2(\omg)$ norm:
\begin{equation*}
\mcJ(\theta)=\mathbb{E}_{u}\left[\vertii{\mcL_{\theta}[u]-f}_{L^2(\omg)}^2\right]\approx\dfrac{1}{K}\sum_{k=1}^K \vertii{\mcL_{\theta}[u_k]-f_k}_{L^2(\omg)}^2.
\end{equation*}
Therefore, learning the surrogate operator $\mcL_{\theta}$ is equivalent to finding optimal parameters $\theta=\{C_m\}$ by minimizing the objective $\mcJ(\theta)$. Here, we stress that NOR aims to learn a surrogate operator for the ground truth operator $\mcL^+$, rather than the displacement field solution $u(x)$ for a single instance loading field $f(x)$. Since our goal is to provide a homogenized surrogate model which can be readily applied in simulation problems with time-dependent loadings, the former setting is more appropriate, which directly approximates the solution operator and finds the solutions for different loadings at different time instants. 
Moreover, approximating the operator also possesses the notable advantages of resolution independence and convergence. 




\subsection{NOR for Metamaterial Homogenization}\label{sec:nor} 

We now seek nonlocal homogenized models for the stress wave propagation problem in a one-dimensional metamaterial, i.e., $p=1$, using NOR as described above. As illustrated in Figure \ref{fig:bar}, each material is a heterogeneous bar formed by two dissimilar materials, with microstructure either made of periodic layers or randomly generated layers. The goal of NOR is to learn a surrogate model which is able to reproduce wave propagation on distances that are much larger than the size of the microstructure, without resolving the microscales.

We assume that there exists a (possibly unknown or nonfeasible) high-fidelity (HF) model that faithfully represents the wave propagation in detailed microstructures: for $(x,t)\in\omg\times[0,T]$,
\begin{equation}
\dfrac{\partial^2 \hat{u}}{\partial t^2}(x,t)-\mcL^+[\hat{u}](x,t)=g(x,t).   
\end{equation}
Here $\mcL^+$ is the HF operator that considers the detailed microstructure, $\hat{u}(x,t)$ is the HF solution which can be provided either from fine-scale simulations or from experiment measurements in practice, and $g(x,t)$ represents a time-dependent force loading term. Analogously, we will refer to the homogenized effective nonlocal operator as $\mcL_{\theta}$, and assume that the surrogate model has the form
\begin{equation}\label{eqn:homo}
\dfrac{\partial^2 \tilde{u}}{\partial t^2}(x,t)-\mcL_{\theta}[\tilde{u}](x,t)=g(x,t),   
\end{equation}
for $(x,t)\in\Omega\times[0,T]$, augmented with Dirichlet-type boundary conditions from the high-fidelity solution on a layer of thickness $\delta$ that surrounds the domain, and the same initial conditions as in the high-fidelity model. Here 
$\tilde{u}(x,t)$ is the homogenized solution. As shown in \cite{du2017peridynamic}, the second-order-in-time nonlocal equation in \eqref{eqn:homo} is guaranteed to be well-posed as far as $\gamma_\theta$ is uniformly Lipschitz continuous. 
Therefore, when parameterizing the nonlocal kernel $\gamma_\theta$ as a linear combination of basis polynomials following \eqref{eqn:linear}, one can make sure that the learnt model can be readily applied in simulation problems. 

To learn the optimal $\mcL_{\theta}$, suppose we have $K$ observations of forcing terms $g_k(x,t)$ and the corresponding high-fidelity solution/experimental measurements $\hat{u}_k(x,t)$, $k=1,\cdots,K$, measured at time instance $t^{n}\in[0,T]$ and discretization points $x_i\in\omg$. Without loss of generality, here we assume measurements are provided on uniformly spacing spatial and time instances, with fixed spatial grid size $\Delta x$ and time step size $\Delta t$. Denoting the collection of discretization points as $\chi=\{x_i\}_{i=1}^L$, then the training dataset contains $N_{train}:=LK\lfloor T/\Delta t\rfloor$ measurements in total, specifically, 
$$\mcD_{train}=\{(\hat{u}_k(x_i,t^n),g_k(x_i,t^n))\}_{k,i,n=1}^{K,L,\lfloor T/\Delta t\rfloor}.$$
In NOR 
the squared loss then writes:
{\begin{align}
\nonumber\mcJ(\theta)\approx&\dfrac{1}{K}\sum_{k=1}^K \vertii{\mcL_{\theta}[\hat{u}_k]-\mcL^+[\hat{u}_k]}_{L^2(\omg\times[0,T])}^2\\
=&\dfrac{1}{K}\sum_{k=1}^K \vertii{\mcL_{\theta}[\hat{u}_k]-\dfrac{\partial^2\hat{u}_k}{\partial t^2}+g_k(x,t)}_{L^2(\omg\times[0,T])}^2.\label{eqn:loss}
\end{align}}
To numerically evaluate the above loss, we discretize $\frac{\partial^2\hat{u}_k}{\partial t^2}$ with the central difference scheme in time and Riemann sum approximation of the nonlocal operator in space: 
\begin{align}
\dfrac{\partial^2\hat{u}_k}{\partial t^2}&(x,t)\approx\ddot{\hat{u}}_k(x,t):=\dfrac{1}{\Delta t^2}(\hat{u}_k(x,t+\Delta t)- 2\hat{u}_k(x,t)+\hat{u}_k(x,t-\Delta t)),\label{eqn:utt}\\
\nonumber\mcL_{\theta}&[\hat{u}_k](x,t)=\int_{B_\delta(x)}\gamma_{\theta}(\verti{y-x})(\hat{u}_k(y,t)-\hat{u}_k(x,t))dy \approx\mcL_{\theta,\Delta x}[\hat{u}_k](x,t)\\
\nonumber:=&\Delta x\sum_{x_j\in B_\delta(x)\cap\chi}\sum_{m=1}^M C_mP_m(\verti{x_j-x}) (\hat{u}_k(x_j,t)-\hat{u}_k(x,t))\\
=&\Delta x\sum_{\alpha=1}^d\sum_{m=1}^M C_mP_m(\verti{x_j-x}) (\hat{u}_k(x+\alpha\Delta x,t)-2\hat{u}_k(x,t)+\hat{u}_k(x-\alpha\Delta x,t)),\label{eqn:Lh}
\end{align}
for each $x=x_i\in\chi$, $t=t^n$ and $d:=\lfloor\delta/\Delta x\rfloor$. Substituting the above schemes into \eqref{eqn:loss}, we then obtain
\begin{align}
\mcJ(\theta)\approx&\dfrac{1}{K}\sum_{k=1}^K\sum_{x_i\in\chi}\sum_{n=1}^{\lfloor T/\Delta t\rfloor} \left({y}_{k,i}^{n}-(\mathbf{s}_{k,i}^{n})^T\mathbf{B}\mathbf{C}\right)^2.\label{eqn:linearr}
\end{align}
where the (reformulated) data pair ${y}_{k,i}^{n}\in \real$ and $\mathbf{s}_{k,i}^{n}\in \real^{d}$ are defined as
\begin{align}
{y}_{k,i}^{n}:=&\dfrac{1}{\Delta t^2}\left(2\hat{u}_k(x_i,t^{n})-\hat{u}_k(x_i,t^{n+1})-\hat{u}_k(x_i,t^{n-1})\right)+g_k(x_i,t^n),\label{eqn:y}\\
[\mathbf{s}_{k,i}^{n}]_{\alpha}:=&\Delta x\left(\hat{u}_k(x_{i+\alpha},t^{n})+\hat{u}_k(x_{i-\alpha},t^{n})-2\hat{u}_k(x_i,t^{n})\right), \;\alpha=1,\cdots,d.\label{eqn:s}
\end{align}
The parameter vector $\mathbf{C}:=[C_1,\cdots,C_M]\in \real^{M}$, and the feature matrix $\mathbf{B}:=(\mathbf{b}_1,\cdots,\mathbf{b}_M)\in \real^{d\times M}$ is  defined as:
$$\mathbf{B}_{\alpha m}=P_m(\alpha\Delta x),\,m=1,\cdots,M,\;\alpha=1,\cdots,d,$$
and the dimension of each feature equal to $d$. Therefore, the optimal parameters are obtained by solving a (constrained) optimization problem
\begin{align}
\underset{\theta}{\text{min}}&\;\mcJ(\theta)+\zeta\mcR(\theta), \quad\text{s.t. }\mcL_{\theta} \text{ satisfies physics-based constraints.} \label{eqn:opt}
\end{align}
Here $\mcR(\theta)$ is a regularization term which aims to prevent over-fitting in the inverse problem, and $\zeta$ is the regularization parameter. A commonly used regularization term is the Euclidean norm in the classical Tikhonov regularization, i.e., $\mcR(\theta):=\vertii{\mathbf{C}}_{l^2}^2$. The physics constraints denotes the additional conditions which enforce partial physical knowledge of the heterogeneous material, for which we will explain later on.

From \eqref{eqn:linearr}, we can see that NOR is equivalent to a linear model with $M$-dimensional features in the kernel space. 
When taking $P_m$, $m=1,\cdots,d=M$ as the Lagrange basis polynomials, satisfying $P_m(\alpha\Delta x)=1$ for $\alpha=m$ and $P_m(\alpha\Delta x)=1$ for all $\alpha\neq m$, we note that $\mathbf{B}$ becomes an identity matrix and NOR will be equivalent to a linear regression problem. Therefore, algorithms and analysis on linear regression models can be immediately applied in NOR. In empirical experiments of Section \ref{sec:exp}, this linear kernel regression (LR) setting will be employed as the baseline method on a new microstructure (task), which only uses data generated from that task.

\section{ Proposed Meta-Learning Algorithm}

We now consider the problem of meta-learning in NOR, such that multiple tasks share a common set of low-dimensional features in the kernel space. Given $K_{train}$ observations $(\hat{u}_k(x,t),g_k(x,t))$ which belong to $H$ unobserved underlying tasks, the meta-learning NOR model writes:
\begin{equation}\label{eqn:metamodel_full}
\dfrac{\partial^2\hat{u}_{k}}{\partial t^2}(x,t)-\mcL_{\theta_{\eta(k)}}[\hat{u}_k](x,t)=g_k(x,t)+\epsilon_k(x,t).
\end{equation}
Here, we assume that each task corresponds to a different microstructure and the underlying optimal surrogate kernel $\gamma_{\theta_{\eta(k)}}$, where $\eta(k)\in\{1,\cdots,H\}$ denotes the index of task associated with the $k$-th function pair. $\epsilon_{k}(x,t)$ is a smooth function related with the additive noise, describing the discrepancy between the ground-truth operator and the optimal surrogate operator, i.e., $\epsilon_{k}(x,t):=(\mcL^+-\mcL_{\theta_{\eta(k)}})[\hat{u}_k](x,t)$. Our goal is to recover the underlining low dimensional representation for the kernel space, and use this representation to recover a better estimate for a new and unseen task. Mathematically, we assume that there exists an unobserved kernel feature space $span(\{P_m(\verti{y-x})\}_{m=1}^M)$, such that $M\ll d$ and
\begin{equation}\label{eqn:metakernel}
\gamma_{\theta_\eta}(\verti{y-x})=\sum_{m=1}^M \mathbf{C}_{\eta,m}P_m(\verti{y-x}), 
\end{equation}
where $\mathbf{C}_{\eta}$ is the parameter for the $\eta-$th task. For a new and unseen microstructure, we assume there also exists a surrogate model for it:
\begin{equation}\label{eqn:metamodel_test}
\dfrac{\partial^2\hat{u}}{\partial t^2}(x,t)-\mcL_{\theta_{test}}[\hat{u}](x,t)=g(x,t)+\epsilon(x,t),
\end{equation}
with its optimal surrogate kernel inside the unobserved kernel feature space $span(\{P_m(\verti{y-x})\}_{m=1}^M)$, i.e., there exists $\mathbf{C}_{test}\in\real^M$ such that
\begin{equation}\label{eqn:metatestkernel}
\gamma_{\theta_{test}}(\verti{y-x})=\sum_{m=1}^M \mathbf{C}_{test,m}P_m(\verti{y-x}). 
\end{equation}
Then, with only a few measurements given for a new and unseen microstructure, we aim to efficiently recover this optimal estimator $\gamma_{\theta_{test}}$.

Comparing with NOR and the other classical machine learning methods, our method aims at five desirable properties: 1) The learning algorithm is sample efficient on the new task, which implies that the optimal estimator $\gamma_{\theta_{test}}$ can be learnt even with very scarce measurements. 2) The estimator is resolution independent, in the sense that the learnt model can be applied to different resolution problems. 3) Beyond resolution invariance, we further aim for a robust consistent estimator, that is, the estimator converges as data resolution refines. 4) The method learns the nonlocal surrogate model directly from data, i.e., no preliminary knowledge on the governing law is required. 5) The learnt model is generalizable, meaning that it is applicable to problem settings that are substantially different from the ones used for training in terms of loading and domain/time scales. Hence, once the nonlocal surrogate model is learnt, one can further employ it in further prediction tasks with a longer simulation time, a larger computational domain, and on a different grid.


Before demonstrating our main algorithm, we first establish the connection of our meta-learning model with the linear representation model illustrated in \cite{tripuraneni2021provable}. Following a similar derivation in Section \ref{sec:nor}, we reformulate the training data following \eqref{eqn:y} and \eqref{eqn:s}, then denote the collection of all (reformulated) data points as
$${\mcD}_{train}=\{(y_{k,i}^n,\mathbf{s}^n_{k,i})\}_{k,i,n=1}^{K_{train},L_{train},\lfloor T_{train}/\Delta t\rfloor}=\{(y_j,\mathbf{s}_j)\}_{j=1}^{N_{train}}.$$
With a slight abuse of notations, in the meta-learning algorithm we consider a uniform task sampling model which does not differentiate the datapoints from different sample $k$, time step $n$ and spatial grid $i$, then use ${\eta}(j)\in\{1,\cdots, H\}$ to denote the index of the task associated with the datapoint $j$. Then, discovering the kernel basis $\{P_m(\verti{y-x})\}_{m=1}^M$ is equivalent to recovering a linear feature matrix $\mathbf{B}\in\real^{d\times M}$ with orthonormal columns, such that $P_{m}(\alpha\Delta x)=\mathbf{B}_{\alpha m}$ and
\begin{equation}\label{eqn:metamodel}
{y}_{j}=\mathbf{s}_j^T\mathbf{B}\mathbf{C}_{\eta(j)}+\epsilon_j,
\end{equation}
where $\mathbf{C}_{\eta(j)}$ is the parameter for the $\eta(j)-$th task, and $\epsilon_{j}$ is additive noise. For the new task, with a collection of (reformulated) datapoints
$${\mcD}_{test}=\{(y^{test}_j,\mathbf{s}^{test}_j)\}_{j=1}^{N_{test}},$$
recovering the optimal kernel $\gamma_{\theta_{test}}$ is equivalent to recovering an optimal estimate $\mathbf{C}_{test}$, such that:
\begin{equation}\label{eqn:metatest}
{y}^{test}_{j}=(\mathbf{s}^{test}_j)^T\mathbf{B}\mathbf{C}_{test}+\epsilon^{test}_{j}. 
\end{equation}

Our meta-learning model has two stages: firstly, the linear feature matrix $\mathbf{B}$ is recovered from $\mathcal{D}_{train}$, the data from the first $H$ known tasks, then the learnt feature representation will be employed to discover an estimate of the task parameter $\mathbf{C}_{test}^\circ$ from a (scarce) test dataset corresponding to this new task. In particular, we employ the provable meta-learning algorithm recently proposed in \cite{tripuraneni2021provable}. In the following we briefly describe the main steps, with further theoretical results elaborated in Section \ref{sec:error} and discussions on physical constraints in Section \ref{sec:physics}. 

\textbf{Step 1: Data Preprocessing for Learning Tasks.} 
We first normalize each pair of solution $\hat{u}_k(x,t)$ and forcing term $g_k(x,t)$ with respect to the $L^2$ norm of $\hat{u}_k(x,t)$, then generate the reformulated data pairs following \eqref{eqn:y} and \eqref{eqn:s}. To further ensure that the dataset satisfies the sub-Gaussian requirement (see Assumption \ref{asp:2}), we normalize the training data pairs such that $\mathbb{E}[\mathbf{s}]=\mathbf{0}$ and $\mathbb{E}[\mathbf{s}\mathbf{s}^T]=\mathbf{I}_d$.

\textbf{Step 2: Meta-Train to Learn Kernel Features.} As the first stage of meta-learning, we solve for $\mathbf{B}\in\real^{d\times M}$ and try to recover $\mathbf{W}:=(\mathbf{C}_1,\cdots,\mathbf{C}_H)^T\mathbf{B}^T\in\real^{H\times d}$ with $rank(\mathbf{W})=r<d$. In particular, we consider the Burer-Monteiro factorization of $\mathbf{W}=\mathbf{U}\mathbf{V}^T$ with $\mathbf{U}\in\real^{H\times M}$, $\mathbf{V}\in\real^{d\times M}$, and solve the following optimization problem
\begin{align}
\underset{\mathbf{U},\mathbf{V}}{\min}&\,\dfrac{H}{N_{train}}\sum_{j=1}^{N_{train}}\left({y}_{j}-\mathbf{e}_{\eta(j)}\cdot\left[\mathbf{s}_{j}^T\mathbf{V}\mathbf{U}^T\right]\right)^2+\frac{1}{4}\vertii{\mathbf{U}^T\mathbf{U}-\mathbf{V}^T\mathbf{V}}_F^2,\label{eqn:FO}
\end{align}
where $N_{train}=K_{train}L_{train}\lfloor T_{train}/\Delta t\rfloor$ is the total number of training datapoints and $\mathbf{e}_{j}$ is the $j-$th standard basis vector in $\real^M$. The estimated feature matrix $\mathbf{B}^\circ$ can then be extracted, as an orthonomal basis from the column space of $\mathbf{V}^\circ$. As shown in \cite{ge2017no}, all local minima of this optimization problem, $\mathbf{V}^\circ$, would be in the neighborhood of the optimal, that means, the approximated basis $\mathbf{B}^\circ$ would provide a good estimate to the optimal low-dimensional feature space. 



\textbf{Step 3: Meta-Test to Transfer Features to New Tasks.} As the second stage of meta-learning, we substitute the learnt feature matrix $\mathbf{B}^\circ$ from the first stage, and estimate the new task parameter $\mathbf{C}_{test}$ as follows:
\begin{equation}
 \mathbf{C}^\circ_{test}=\underset{\mathbf{C}_{test}}{\text{argmin}} \sum_{j=1}^{N_{test}}\left(y_j^{test}-(\mathbf{s}_j^{test})^T\mathbf{B}^\circ \mathbf{C}_{test}\right)^2.
\end{equation}
Since this is an ordinary least-square objective, an analytical solution can be obtained:
\begin{equation}\label{eqn:test}
\mathbf{C}^\circ_{test}=\left(\sum_{j=1}^{N_{test}}(\mathbf{B}^\circ)^T  \mathbf{s}_j^{test}(\mathbf{s}_j^{test})^T\mathbf{B}^\circ\right)^\dag (\mathbf{B}^\circ)^T \sum_{j=1}^{S_{test}} \mathbf{s}_j^{test} y_j^{test}
\end{equation}
where $\dag$ indicates the Moore-Penrose pseudo-inverse. 

\textbf{Step 4: Postprocessing to Obtain a Continuous Model.} To construct the continuous model which can be employed in further prediction tasks with various resolutions, we employ the B-spline basis functions consisting of piece-wise polynomials with degree 2. In particular, we construct the basis polynomials as $P^\circ_m(\verti{y-x}):=\sum_{\alpha=1}^d\mathbf{B}^\circ_{\alpha m} N_{\alpha,2}(\verti{y-x})$, $m=1,\cdots,M$, where $N_{\alpha,2}$ are constructed with evenly spaced knots on interval $[0,(d+1)\Delta x]$. Substituting the chosen polynomials $\{P_m(\verti{y-x})\}$ into \eqref{eqn:linear}, we obtain the learnt nonlocal surrogate model for the new microstructure, which is defined by a continuous nonlocal kernel
\begin{equation}
\gamma^\circ_{\theta_{test}}(\verti{y-x}):=\sum_{m=1}^M(\mathbf{C}^\circ_{test})_m P^\circ_m(\verti{y-x}).
\end{equation}
Note that this kernel is indeed a twice differentiable function for $\verti{y-x}\leq \delta$, the resultant nonlocal surrogate model is therefore well-posed and defined in a continuous way.



\subsection{Prediction Error Bounds}\label{sec:error}

We now provide error bounds for MetaNOR based on the results for linear regression provided in \cite{tripuraneni2021provable}. Throughout this section, we use $\vertii{\vb}_{l^2(\omg)}:=\sqrt{\Delta x \sum_{x_i\in\chi}v_i^2}$ to denote the domain-associated $l^2$ norm for a vector with values on $\chi$. This norm can be seem as a discretized approximation for the $L^2(\omg)$ norm. Consider a solution pair $(\Cb^\circ,\Bb^\circ)$ which corresponds to a local minimizer of \eqref{eqn:FO} and \eqref{eqn:test}, we use $\mcL^\circ_{\theta_{test}}$ to denote the corresponding nonlocal operator generated from the learnt nonlocal kernel $\gamma^\circ_{\theta_{test}}$, and $\mcL^\circ_{\theta_{test},h}$ to denote its approximation by Riemann sum following \eqref{eqn:Lh}.  

We now provide the error estimates for the kernel estimator, $\vertii{\gamma_{test}-\gamma^\circ_{\theta_{test}}}_{l^2([0,\delta])}$, and for the prediction error. For the later, we consider a given time-dependent loading $g(x,t)$ with $x\in\Omega_{pred}$ and $t\in [0,T_{pred}]$, then use the learnt nonlocal model to predict the material response of our test microstructure, i.e., to provide an approximated displacement solution. Here, we stress that the prediction domain $\omg_{pred}$ and time interval $[0,T_{pred}]$ may be different from the training datasets. We then discretize $\omg_{pred}$ and $[0,T_{pred}]$ with grid sizes $\Delta x$ and $\Delta t$, respectively, and denote the spatial grid set as $\chi_{pred}$. For simplicity of analysis, here we take the same discretization sizes as those in the training dataset. However, since a continuous model is learnt, in practice one may employ different resolution or even discretization methods, as will be numerically demonstrated in the empirical experiment of Section \ref{sec:veri}. We consider $\delta$ as a physical parameter, i.e., as a fixed value, and hence $d=\Theta(\Delta x^{-1})$. Denoting $\hat{u}(\cdot,t^{n})$ as the ground-truth high-fidelity solution subject to loading $g(x,t)$ and $\bar{u}^n(x_i)$ as the numerical solution satisfying
\begin{equation}\label{eqn:modelmeta}
\ddot{\bar{u}}^n(x_i)-\mcL^\circ_{\theta_{\theta_{test}},h}[\bar{u}^n](x_i)=g(x_i,t^n)    
\end{equation}
for $x_i\in\chi_{pred}$ and $n=1,\cdots, \lfloor T_{pred}/\Delta t\rfloor$, we aim to provide the error bound in the discretized energy norm for displacement prediction:
\begin{align*}
\vertii{\bar{u}^n-\hat{u}(\cdot,t^{n})}_{E(\omg)}:=&\sum_{i=1}^{L_{pred}}\left(\dfrac{e_i^{n}-e_i^{n-1}}{\Delta t}\right)^2\\
&+2\Delta x\sum_{\alpha=1}^d\sum_{i=1-\alpha}^{L_{pred}} (\mathbf{B}^\circ\mathbf{C}_{test}^\circ)_\alpha(e^{n}_{i+\alpha}-e_i^{n})^2,\\
\text{where } e_i^n:=&\hat{u}(x_i,t^n)-\bar{u}^n(x_i).
\end{align*}

We first detail three required assumptions for the analysis. In the following derivations, we always assume that the statements below are true, and therefore will not list them in the statement of theorems again.

\begin{asp}[Sub-Gaussian Design and Noise]\label{asp:2}
For both the training and test datasets, the vectors $\mathbf{s}_j$ are i.i.d. designed with zero mean, covariance $\mathbb{E}[\mathbf{s}\mathbf{s}^T]=\mathbf{I}_d$, and are $\mathbf{I}$-sub-Gaussian. The additive noise variables $\epsilon_j=y_j-\mathbf{s}^T_j\mathbf{B}\mathbf{C}_{{\eta}(j)}$ are also i.i.d. sub-Gaussian with variance parameter $1$. Moreover, $\epsilon_j$ are independent of $\mathbf{s}_j$.
\end{asp}

For the $H$ training tasks, we define the population task diversity matrix and condition numbers as $\mathbf{T}=(\mathbf{C}_1,\cdots,\mathbf{C}_H)^T\in\real^{H\times M}$, $\nu:=\sigma_M(\mathbf{T}^T\mathbf{T}/H)$ and $\kappa:=\frac{1}{\nu}\sigma_1(\mathbf{T}^T\mathbf{T}/H)$.
\begin{asp}[Task-Diversity and Normalization]\label{asp:3}
The $H$ underlying task parameters $\mathbf{C}^*_j$ satisfies $\vertii{\mathbf{C}_j}_{l^2}=\Theta(1)$, i.e., they are asymptotically bounded below and above by constants. Moreover, the population task diversity matrix is well-conditioned, i.e., $\kappa\leq O(1)$, which indicates that 
$\nu\geq\omg(1/M)$.
\end{asp}

Moreover, we make the following additional assumptions associated with the stability and consistency of the numerical scheme:
\begin{asp}[Numerical Stability and Model Consistency]\label{asp:stab}
The high-fidelity solution for our prediction task $\hat{u}\in C^2(\omg\times[0,T_{pred}])$ and $\Delta t$ is sufficiently small such that it satisfies $\Delta t\leq \min[(8\Delta x\vertii{\Bb^\circ\Cb^\circ}_{l^1})^{-1},(2T_{pred})^{-1}]$.
{Moreover, the modeling error, $\epsilon(x,t)$, is bounded by a constant $E$ for all $x\in\omg_{pred}$ and $t\in[0,T_{pred}]$.}
\end{asp}



We now proceed to provide error bounds to our linear kernel representation learning setting. 

\begin{thm}\label{thm:main}
Suppose we are given $N_{train}$ total training datapoints from $H$ diverse and normalized tasks, and $N_{test}$ numbers of test datapoint on a new task with unknown microstructure. If the number of meta-train samples $N_{train}$ satisfies $N_{train}\gtrsim \text{polylog}(N_{train},d,H)(\kappa M)^4\max\{H,d\}$, the number of meta-test samples $N_{test}$ satisfies $N_{test}\gtrsim M \log(N_{test})$, and the optimal test microstructure satisfies $\vertii{\Cb^*_{test}}_{l^2}\leq O(1)$, then any local minimizer of \eqref{eqn:FO} and the learnt kernel converges to the underlying optimal kernel with the following error bound:
$$\vertii{\gamma_{test}-\gamma^\circ_{\theta_{test}}}^2_{l^2([0,\delta])}\leq \tilde{O}\left(\Delta x\dfrac{\max\{H,d\} M^2}{N_{train}}+\Delta x\dfrac{M}{N_{test}}\right),$$
and the corresponding approximated solution $\bar{u}$ has the following excess prediction error bound for $n=1,\cdots,\lfloor T_{pred}/\Delta t\rfloor$:
$$\vertii{\bar{u}^{n}-\hat{u}(\cdot,t^{n})}_{E(\omg)}^2\leq \tilde{O}\left(E^2+\Delta t^4+\Delta x^2+\Delta x\left[\dfrac{\max(H,d)M^2}{N_{train}}+\dfrac{M}{N_{test}}\right]\right),$$
with probability at least $1-O((\text{poly}(d))^{-1}+N_{test}^{-100})$.
\end{thm}


The proof is obtained by applying Theorems 2 and 4 in \cite{tripuraneni2021provable}. 
A more detailed proof is provided in Appendix. 

\begin{remark}
This theorem indicates that when a sufficiently large training dataset is provided, for a new microstructure with very scarce measurements ($N_{test}\ll N_{train}/(\max(H,d)M)$), we have an approximated kernel error bound as $\tilde{O}\left(\left(\frac{M\Delta x}{N_{test}}\right)^{1/2}\right)$ and the energy error bound for the prediction task as $\tilde{O}\left(E+(\Delta t)^2+\left(\frac{M\Delta x}{N_{test}}\right)^{1/2}\right)$. Hence, when the nonlocal model serves as a good surrogate for the material response, i.e., $E$ is negligible, the estimator from MetaNOR provides a converging kernel and solution for further prediction tasks.
\end{remark}

\subsection{Physics-Based Constraints}\label{sec:physics}

As illustrated in \cite{You2021}, when some physical knowledge is available, 
these knowledge can be incorporated into the optimization problem as physics-based constraints in \eqref{eqn:opt}. In particular, when the effective wave speed for infinitely long wavelengths, $c_0$, is available, the corresponding constraint writes:
\begin{equation}\label{eqn:cond1}
    \int_{0}^\delta \xi^2\gamma_\theta(\verti{\xi})d\xi=\bar{\rho} c_0^2,
\end{equation}
where $\bar{\rho}$ is the effective material density. Discretizing \eqref{eqn:cond1} by Riemann sum, we obtain the first constraint of $\{C_m\}$:
\begin{equation}\label{eqn:cond1_disc}
    \bar{\rho} c_0^2=\sum_{m=1}^{M} C_m\sum_{\alpha=1}^d \alpha^2 \Delta x^3 P_m(\alpha\Delta x)=\sum_{m=1}^{r} C_m A_{1m}
\end{equation}
where $A_{1m}:=\sum_{\alpha=1}^d \alpha^2 \Delta x^3 P_m(\alpha\Delta x)$. Furthermore, when the curvature of the dispersion curve in the low-frequency limit, $R$, is also available, the corresponding constraint writes:
\begin{equation}\label{eqn:cond2}
    \int_{0}^\delta \xi^4 \gamma_\theta(\verti{\xi})d\xi=-4\bar{\rho} c_0^3R.
\end{equation}
Discretizing \eqref{eqn:cond2} yields the second constraint of $\{C_m\}$:
\begin{equation}\label{eqn:cond2_disc}
    -4\bar{\rho} c_0^3 R=\sum_{m=1}^{M} C_m\sum_{\alpha=1}^d \alpha^4 \Delta x^5 P_m(\alpha\Delta x)=\sum_{m=1}^{M} C_m A_{2m}
\end{equation}
where $A_{2m}:=\sum_{\alpha=1}^d \alpha^4 \Delta x^5 P_m(\alpha\Delta x)$. Therefore, these two physics-based constraints are imposed as linear constraints for $\{C_m\}$. In empirical tests, we will refer to the experiments with these constraints applied as the ``constraint'' cases. In this work, we consider the heterogeneous bar composed by alternating layers of two dissimilar materials, with (averaged) layer size $L_1=(1-\phi)L$, $L_2=(1+\phi)L$ for components 1 and 2, respectively. Then the effective material density, Young's modulus, the wave speed, and the dispersion curvature are given by $\bar{\rho}=((1-\phi)\rho_1+(1+\phi)\rho_2)/2$, $\bar{E}=2/((1-\phi)E_1^{-1}+(1+\phi)E_2^{-1})$,  $c_0=\sqrt{\bar{E}/\bar{\rho}}$, and $R=0$. 
To apply \eqref{eqn:cond1_disc} and \eqref{eqn:cond2_disc}, we reformulate the constraint optimization problem such that an unconstraint optimization problem of the form \eqref{eqn:linearr} is obtained. Detailed derivation is provided in the Appendix.

\section{Empirical Experiments}\label{sec:exp}

We evaluate MetaNOR on both synthetic and real-world datasets. On each dataset, we compare our {\meta} approach with  baseline NOR in \eqref{eqn:opt}. For NOR, we use linear regression with the L-curve method to select the proper regularization parameter $\zeta$. In synthetic datasets, we generate the data from a known nonlocal diffusion equation and study the convergence of estimators to the true kernel. We also apply our method to a real-world dataset for stress wave propagation in 1D metamaterials.

In meta-training, we solve the optimization problem \eqref{eqn:FO} using SciPy's L-BFGS-B optimization module. The maximum iteration step is set to $10000$. In meta-testing, the solutions are solved with NumPy's linalg module. 

\subsection{Verification on Synthetic Datasets}\label{sec:veri}

\begin{figure}[h]
\centering
\includegraphics[width=1.0\columnwidth]{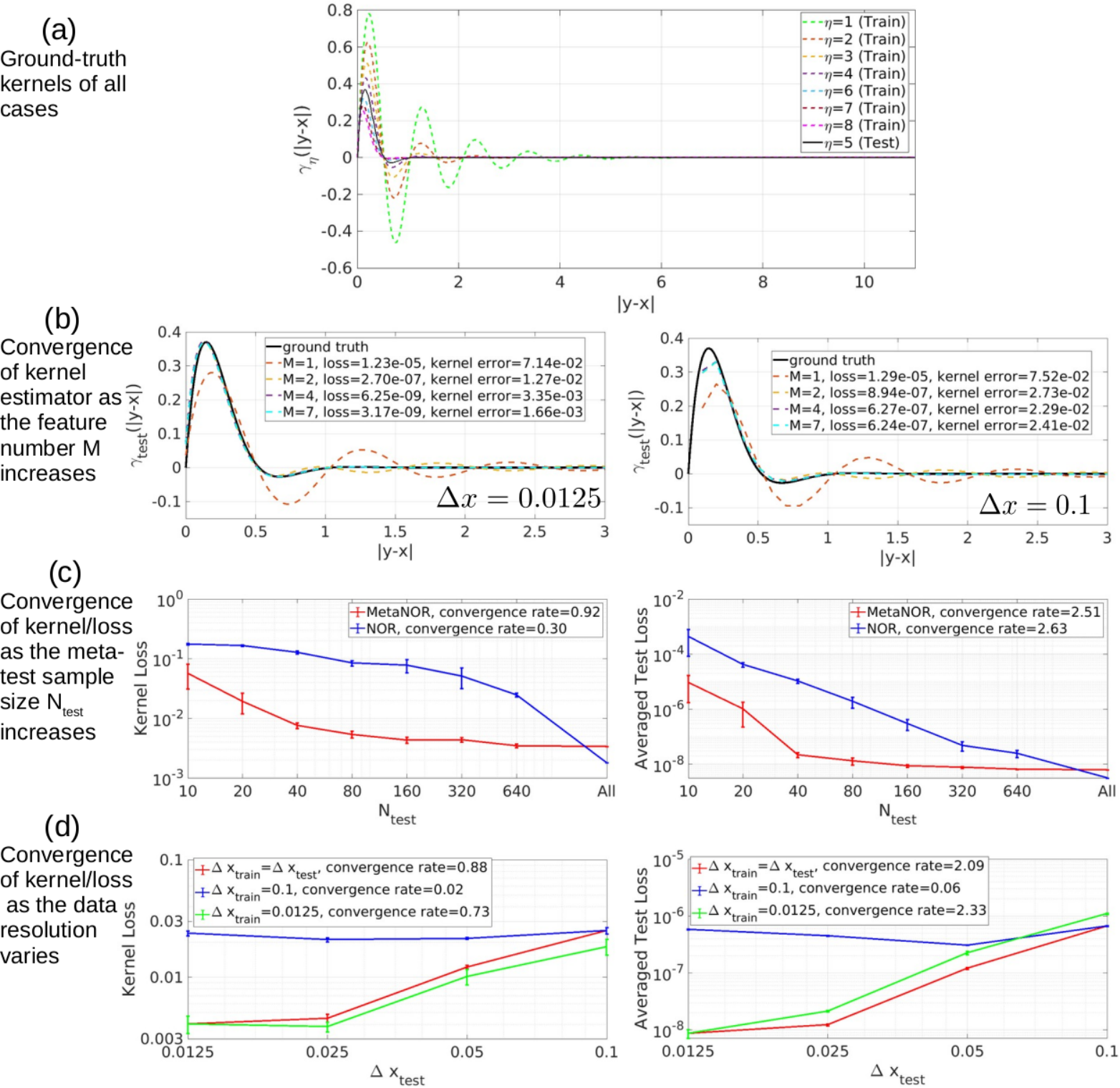}
\caption{Problem settings and convergence study results for the MetaNOR verification on synthetic datasets.}
\label{fig:veri}
\end{figure}

We consider a synthetic dataset generated from a nonlocal diffusion equation 
$$\mcL_{\gamma^+_\eta}[u_k](x):=\int_{B_\delta(x)}\gamma^+_{\eta}(\verti{y-x})(u_k(y)-u_k(x))dy=g_k(x).$$
Here, $\eta$ denotes the index of task, with $\eta\in \{1,\cdots,8\}$. Each task is associated with a sine-type kernel:
$$\gamma^+_{\eta}(\verti{y-x}):=\exp(-\eta(\verti{y-x}))\sin(6\verti{y-x})\mathbf{1}_{[0,10]}(\verti{y-x}),$$
with the estimated support of kernel as $\delta =11$. To generate the training and test function pairs $(u_k(x),g_k(x))$, for each task the kernel acts on the same set of function $\{u_k\}_{k=1,2}$ with $u_1(x)= \sin(x)\mathbf{1}_{[-\pi,\pi]}(x)$ and $u_2(x) =\cos(x)\mathbf{1}_{[-\pi,\pi]}(x)$, and the loading function $\mcL_{\gamma^+_\eta}[u_k]=g_k$ is computed by the adaptive Gauss-Kronrod quadrature method, both on the computational domain $\omg=[-40,40]$. 
To create discrete datasets with different resolutions, we consider $\Delta x\in 0.0125\times\{1,2,4,8\}$. In meta-training, we use all samples from $7$ ``known tasks'' $\eta\in\{1,2,3,4,6,7,8\}$. 
Then, the goal is to learn a good estimator for the ``unknown'' new task with $\eta=5$.

\textbf{Effect of low-dimensional feature selection: }
In this experiment we aim to verify the low-dimensional structure of the kernel space and select a proper value of $M$, with all test measurements employed, i.e., $N_{test}=2\times80/\Delta x$. In Figure \ref{fig:veri}(b) we demonstrate the learnt kernel for $M\in\{1,2,4,7\}$ and $\Delta x\in\{0.0125,0.1\}$, together with the averaged loss on all test samples (denoted as ``loss'') and the $l^2([0,\delta])$ errors for the kernel (denoted as ''kernel error''). It is observed that 
the learnt kernel is visually consistent with the true kernel when $M\geq4$.
Hence in the following investigations we fix $M=4$ for all cases.

\textbf{Sampling efficiency on the new task: } We now demonstrate the performance of the estimator in the small test measurement regime. We randomly select $N_{test}\in\{10,20,40,80,160,320\}$ measurements from all available data on the test task, 
and study the convergence of the learnt kernel as $N_{test}$ increases. In this experiment we fix $M=4$ and $\Delta x=0.0125$. 
To generate a fair comparison, 
the means and standard errors are calculated from 10 independent simulations. The errors of learnt kernels and the averaged loss on all test samples from MetaNOR and NOR are reported in Figure \ref{fig:veri}(c). Averaged  convergence rates are calculated on the relatively small data regime, i.e., for $N_{test}\leq 160$. One can see that the kernel error from MetaNOR decreases almost linearly with the increase of $N_{test}$,
-- a half order faster than the bound suggested in Theorem \ref{thm:main}. This fact indicates a possible improvement of the analysis in the future work. On the other hand, NOR exhibits a much larger error and test loss 
in the same small test data regime, highlighting the advantage of our MetaNOR in sample efficiency.

\textbf{Resolution independence and convergence: } We now study the performance of the estimator in terms of its convergence as the data mesh refines. Two types of experiments are designed, both with limited measurements ($N_{test}=320$). First, we keep the same resolutions ($\Delta x$) in all tasks, to study the convergence of estimators to the true kernel as $\Delta x$ decreases. Additionally, to verify the consistency of estimators across different resolutions, we further investigate their performances when the training tasks and test tasks have different resolutions. In Figure \ref{fig:veri}(d) the kernel errors and test losses are reported, as functions of $\Delta x$ in the test case (denoted as $\Delta x_{test}$). For the first study, a $0.88$ order convergence is observed, which is consistent with the error bound from Theorem \ref{thm:main}. For the second study, one can see that no matter if we extract the features from a relatively coarse grid ($\Delta x_{train}=0.1$) or a fine grid ($\Delta x_{train}=0.0125$), the resultant estimator on the test task pertains a similar accuracy or even achieves convergence as the test grid size $\Delta x_{test}$ refines, when learning from fine measurements. These results highlight the advantage of our method on learning the kernel and the corresponding continuous nonlocal operator instead of learning the solution: the resultant model is not tied to the input's resolution.

\subsection{Application to Wave Propagation in Metamaterials}\label{sec:app}

\begin{figure}[h]
\centering
\includegraphics[width=1.0\columnwidth]{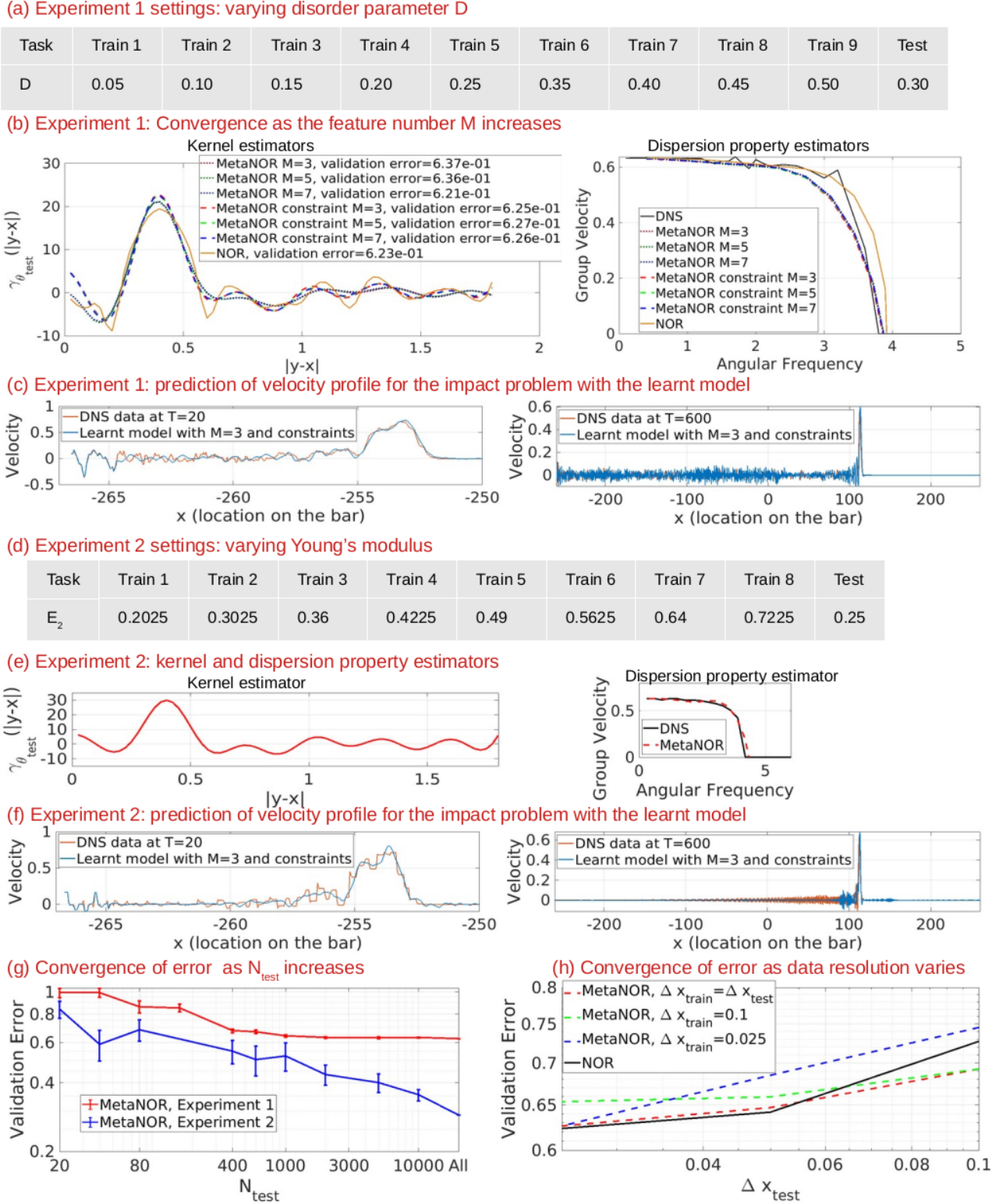}
\caption{Problem settings and numerical results for the MetaNOR application to wave propagation modeling problem in 1D metamaterials.}
\label{fig:gv}
\end{figure}

We now apply MetaNOR to model the propagation of stress waves in one-dimensional metamaterials. Two experiments are considered:
\begin{enumerate}
    \item \textit{(Varying Disorder Parameter, see Figure \ref{fig:gv}(a))} We aim to transfer the knowledge between different disordered microstructures, where the size of each layer is defined by a random variable. For component 1, the layer size $L_1\sim \mathcal{U}[(1-D)(1-\phi)L,(1+D)(1-\phi)L]$, and for component 2 the layer size $L_2\sim \mathcal{U}[(1-D)(1+\phi)L,(1+D)(1+\phi)L]$. Here $D\in[0.05,0.5]$ is the disorder parameter for each task, and the Young's modulus $E_1=1$ and $E_2=0.25$ are fixed. For this experiment we train with $9$ microstructures, then test the meta-learned parameter on a new microstructure with $D=0.3$.
    \item \textit{(Varying Young's Modulus, see Figure \ref{fig:gv}(d))} We aim to transfer the knowledge between varying components. Periodic layers are considered ($D=0$) with fixed Young's modulus $E_1=1$ in component 1 and varying Young's modulus ($E_2\in[0.2025,0.7225]$) in component 2. For this experiment we train with $8$ microstructures, then test the meta-learned parameter on an new microstructure with $E_2=0.25$.
\end{enumerate}
The high-fidelity dataset we rely on is generated by a classical wave solver, where all material interfaces are treated explicitly, and therefore small time and step discretization sizes are required. This solver and its results will be referred to as Direct Numerical Solution (DNS). 
In all training data, we set $L=0.2$, $\rho_1=\rho_2=1$, $\phi=0$, and the domain $\omg=[-50,50]$. Following the settings in \cite{You2021}, two types of data, including $20$ simulations from the oscillating source dataset and $20$ simulations from the plane wave dataset, are generated for meta-training and meta-test in each task. Parameters for the training and the optimization algorithm are set to $\Delta x \in \{0.025,0.05,0.1\}$, $\Delta t = 0.02$, $T= 2$ and $\delta =1.8$. Additionally, we create two validation datasets, denoted as the wave packet dataset and the projectile impact dataset respectively, both very different from the meta-training and meta-test datasets. They consider a much longer bar ($\omg_{wp}=[-133.3, 133.3]$ for wave packet and $\omg_{impact}=[-267,267]$ for impact), under a different loading condition from the training dataset, and with a much longer simulation time ($T_{wp}=100$ and $T_{impact}=600$). 
Full details are provided in Appendix.



\paragraph{Comparison Metrics} Notice that in this case, the data is not faithful to the nonlocal model, but generated from a high-fidelity (HF) model with microscale details. Therefore, there is no ground-truth kernel and we demonstrate the performance of estimators by studying their capability of reproducing the dispersion relation and the wave motion on the two validation datasets, and compare them with the results computed with DNS. The dispersion curve provides the group velocity profile as a function of frequency for each microstructure, which directly depicts the dispersion properties in this microstructure. We further report the prediction error in the discrete energy norm, $\vertii{\bar{u}^{n}-\hat{u}(\cdot,t^{n})}_{E(\omg)}$, on the wave packet dataset. Last, we use the learnt kernel to perform long-term prediction tasks on the projectile impact dataset, to validate the model stability and generalizability.

\paragraph{Model Validation} To investigate the low-dimensional structure of kernel space, in Figure \ref{fig:gv}(b) we report the estimated kernels in experiment 1, their corresponding group velocities, and validation errors for $\Delta x=0.025$ and $M\in\{3,5,7\}$. We can observe that, while all MetaNOR models have successfully reproduces the DNS dispersion relation, the ``constraint'' cases have achieved a better prediction accuracy comparing with the ones without physical constraints. Hence, in the following studies we mainly focus on ``constraint'' cases. We employ the constraint model with $M=3$, $\Delta x=0.05$, to predict the short-term ($T=20$) and long-term ($T=600$) velocity profiles subject to projectile impact, and report the results in Figure \ref{fig:gv}(c). The results are consistent with DNS simulations, verifying that our optimal kernel can accurately predict the short-and long-time wave propagation. We then perform similar tests in experiment 2, with the predicted kernel, dispersion relation, and wave propagation prediction results provided in Figures \ref{fig:gv}(e) and (f). All these results indicate that there exists a common set of low-dimensional features for all microstructures, and MetaNOR provides a good surrogate model based on these features in the low-dimensional kernel space.

\paragraph{Sample Efficiency} We now consider both experiment settings with $M=3$ and $\Delta x=0.025$, and randomly pick $N_{test}\in[10,10^4]$ numbers of datapoints on the new and unseen microstructure. For each $N_{test}$ we repeat the experiment for $10$ times, to plot the mean and standard error of results. Note that under this scarce sample setting the estimated model from NOR gets unstable and fails the prediction task. Hence we only report the MetaNOR results. From Figure \ref{fig:gv}(g) we can see that as the number of test sample increases, for both experiments the validation error decreases. Notice that because of the unavoidable modeling error due to the discrepancy between nonlocal surrogates and the HF model, as shown in Theorem \ref{thm:main}, one should not expect the prediction error to converge to zero. This result again verifies the robustness of MetaNOR in the small data regime.

\paragraph{Resolution Independence and Convergence} Lastly, we consider experiment setting 1 with $M=7$, to study the performance of the estimator in terms of its convergence as the data mesh refines. In Figure \ref{fig:gv}(h) the validation errors are reported for different combinations of $\Delta x_{train}$ and $\Delta x_{test}$. The learnt estimator again demonstrates an improved accuracy as the data resolution refines.



\section{Conclusion}

We proposed a meta-learning approach for the nonlocal  operator  regression, by taking advantages of a common set of low-dimensional features in a multi-task setting for accurate and efficient adaption to new unseen tasks. Specifically, we reformulate the nonlocal operator regression as a linear kernel regression problem and propose {\meta} as a linear kernel feature learning algorithm with provable guarantees. We apply such a method to metamaterial problems and show the superior transfer capability, showing meta-learning is a promising direction for heterogeneous material discovery. Future work could extend our method to obtain sharper estimates and apply to more general material types.

\backmatter

\bmhead{Supplementary information}

\bmhead{Acknowledgments}

Authors are supported by the National Science Foundation under award DMS 1753031, and the AFOSR grant FA9550-22-1-0197. Portions of this research were conducted on Lehigh University's Research Computing infrastructure partially supported by NSF Award 2019035.

The authors would also like to thank Dr. Stewart Silling for sharing the DNS codes and for the helpful discussions. 

\section*{Declarations}

\textbf{Data Availability} The data that support the findings of this study are available from the corresponding author, Yue Yu, upon reasonable request.

\textbf{Conflicts of Interest} On behalf of all authors, the corresponding author states that there is no conflict of interest. 

\begin{appendices}


\section{Related Works}


\paragraph{Material Discovery} Using machine learning techniques for material discovery is gaining more attention in scientific communities \cite{liu2017materials,lu2018accelerated,butler2018machine,cai2020machine}. They has been applied to materials such as thermoelectric material \cite{iwasaki2019machine}, metallic glasses \cite{ren2018accelerated}, high-entropy ceramics \cite{kaufmann2020discovery}, and so on. Learning models for metamaterials has also gained popularity with recent approaches such as \cite{You2020Regression}.

\paragraph{Meta Learning} Meta-learning seeks to design algorithms that can utilize previous experience to rapidly learn new skills or adapt to new environments. There is a vast literature on papers proposing meta learning \cite{finn2017model} methods, and they have been applied to patient survival analysis \cite{qiu2020meta}, few short image classification \cite{chen2021metadelta}, and natural language processing \cite{yin2020meta}, just to name a few. Recently, provably generalizable algorithms with sharp guarantees in the linear setting are first provided \cite{tripuraneni2021provable}.

\paragraph{Transfer and Meta-Learning for Material Modeling} 
Despite its popularity, few work has studied material discovery under meta or even transfer setting.  \cite{kailkhura2019reliable} proposes a transfer learning technique  to exploit correlation among different material properties to augment the features with predicted material properties to improve the regression performance. \cite{mai2021use} uses an ensemble of model and a meta-model to help discovering candidate water splitting photocatalysts. To the best of our knowledge, our work is the first application of transfer or meta learning to heterogeneous material homogenization and discovery.

\section{Detailed Proof for the Error Bounds}

In this section we review two main lemmas from \cite{tripuraneni2021provable}, which provide a theoretical prediction error bound for the meta-learning of linear representation model as illustrated in \eqref{eqn:metamodel} and \eqref{eqn:metatest}. Then we employ these two lemmas and detailed the proof of Theorem \ref{thm:main}, which provides the error bound for the meta-learning of kernel representations and the resultant prediction tasks.

\begin{lem}\cite[Theorem~2]{tripuraneni2021provable}\label{thm:4} Assume that we are in a uniform task sampling model. If the number of meta-train samples $N_{train}$ satisfies $N_{train}\gtrsim \text{polylog}(N_{train},d,H)(\kappa M)^4\max\{H,d\}$ and given any local minimum of the optimization objective \eqref{eqn:FO}, the column space of $\mathbf{V}^*$, spanned by the orthonormal feature matrix $\mathbf{B}^\circ$ satisfies 
\begin{equation}
\sin\theta(\mathbf{B}^\circ,\mathbf{B})\leq {O}\left(\sqrt{\dfrac{\max\{H,d\}M\log N_{train}}{\nu N_{train}}}\right),
\end{equation}
with probability at least $1-1/\text{poly}(d)$.
\end{lem}

Note that Assumption \ref{asp:3} guarantees that $\nu\geq\Omega(1/M)$ and the above theorem yields
\begin{equation}
\sin\theta(\mathbf{B}^\circ,\mathbf{B})\leq \tilde{O}\left(\sqrt{\dfrac{\max\{H,d\} M^2}{N_{train}}}\right),
\end{equation}
with probability at least $1-1/\text{poly}(d)$.

\begin{lem}\cite[Theorem~4]{tripuraneni2021provable}\label{thm:6} Suppose the parameter associated with the new task satisfies $\vertii{\mathbf{C}_{test}}_{l^2}\leq O(1)$, then if an estimate $\mathbf{B}^{\circ}$ of the true feature matrix $\mathbf{B}$ satisfies $\sin\theta(\mathbf{B}^{\circ},\mathbf{B})\leq \varpi$ and $N_{test}\gtrsim M\log N_{test}$, then the output parameter $\mathbf{C}_{test}^\circ$ from \eqref{eqn:test} satisfies
\begin{equation}
\vertii{\mathbf{B}^{\circ}\mathbf{C}_{test}^\circ-\mathbf{B}\mathbf{C}_{test}}_{l^2}^2\leq \tilde{O}\left(\varpi^2+\dfrac{M}{N_{test}}\right),
\end{equation}
with probability at least $1-O(N_{test}^{-100})$. 
\end{lem}

Combining Lemma \ref{thm:6} with Lemma \ref{thm:4}, we obtain the following result for applying \eqref{eqn:FO} and \eqref{eqn:test} as a linear feature meta-learning algorithm:
\begin{equation}\label{eqn:linearmeta}
\vertii{\mathbf{B}^{\circ}\mathbf{C}_{test}^\circ-\mathbf{B}\mathbf{C}_{test}}_{l^2}^2 \leq \tilde{O}\left(\dfrac{\max\{H,d\} M^2}{N_{train}}+\dfrac{M}{N_{test}}\right),    
\end{equation}
with probability at least $1-O((\text{poly}(d))^{-1}+N_{test}^{-100})$.

We now proceed to provide the proof for Theorem \ref{thm:main}. In the following, we use $C$ to denote a generic constant which is independent of $\Delta x$, $\Delta t$, $M$, $H$, $N_{train}$ and $N_{test}$, but might depend on $\delta$.

\begin{proof}
With \eqref{eqn:linearmeta}, we immediately obtain the $l^2([0,\delta])$ error estimate for the learnt kernel $\gamma^\circ_{test}$ as
\begin{align*}
\vertii{\gamma_{test}-\gamma^\circ_{\theta_{test}}}^2_{l^2([0,\delta])}=&\Delta x\sum_{\alpha=1}^d \left(\sum_{m=1}^M(\Cb^\circ_{test}-\Cb_{test})_m P_m(\verti{\alpha \Delta x})\right)^2\\
=&\Delta x\vertii{\mathbf{B}\mathbf{C}_{test}-\mathbf{B}^\circ\mathbf{C}_{test}^\circ}^2_{l^2}\\
=&\tilde{O}\left(\Delta x\dfrac{\max\{H,d\} M^2}{N_{train}}+\Delta x\dfrac{M}{N_{test}}\right),
\end{align*}
with probability at least $1-O((\text{poly}(d))^{-1}+N_{test}^{-100})$. 

For the error bound in the discretized energy norm, we notice that the ground-truth solution $\hat{u}$ satisfies:
\begin{align*}
\ddot{\hat{u}}(x_i,t^{n})=& \mcL_{\theta_{test},h}[\hat{u}](x_i,t^n)+g(x_i,t^n)+\epsilon(x_i,t^n)+\left[\ddot{\hat{u}}(x_i,t^{n})-\dfrac{\partial^2 \hat{u}}{\partial t^2}(x_i,t^n)\right]\\
&+\left[\mcL_{\theta_{test}}[\hat{u}](x_i,t^n)-\mcL_{\theta_{test},h}[\hat{u}](x_i,t^n)\right]
\end{align*}
for all $x\in \chi_{pred}$, $n=1,\cdots,\lfloor T_{pred}/\Delta t\rfloor$. Subtracting this equation with \eqref{eqn:modelmeta} and denoting $e_i^n:=\hat{u}(x_i,t^n)-\bar{u}^n(x_i)$, we then obtain
\begin{equation}\label{eqn:e1}
\dfrac{e_i^{n+1}-2e_i^{n}+e_i^{n-1}}{\Delta t^2}= \Delta x\sum_{\alpha=1}^d (\mathbf{B}^\circ\mathbf{C}_{test}^\circ)_\alpha (e_{i+\alpha}^{n}+e_{i-\alpha}^{n}-2e_{i}^{n})+(\epsilon_{all})_i^n,
\end{equation}
where
\begin{align*}
(\epsilon_{all})_i^n:=&\epsilon(x_i,t^n)+\Delta x\sum_{\alpha=1}^d (\mathbf{B}\mathbf{C}_{test}-\mathbf{B}^\circ\mathbf{C}_{test}^\circ)_\alpha (\hat{u}(x_{i+\alpha},t^{n})+\hat{u}(x_{i-\alpha},t^{n})-2\hat{u}(x_{i},t^{n}))\\
&+\left[\ddot{\hat{u}}(x_i,t^{n})-\dfrac{\partial^2 \hat{u}}{\partial t^2}(x_i,t^n)\right]+\left[\mcL_{\theta_{test}}[\hat{u}](x_i,t^n)-\mcL_{\theta_{test},h}[\hat{u}](x_i,t^n)\right].
\end{align*}
With Assumption \ref{asp:stab}, we have the truncation error for the Riemann sum part as $\verti{\mcL_{\theta_{test}}\hat{u}(x,t) -\mcL_{\theta_{test},h}\hat{u}(x,t)}\leq C\Delta x$ for a constant $C$ independent of $\Delta x$ and $\Delta t$ but might depends on $\delta$. Similarly, we have the truncation error for the central difference scheme as $\verti{\ddot{\hat{u}}(x_i,t^{n})-\dfrac{\partial^2 \hat{u}}{\partial t^2}(x_i,t^n)}\leq C(\Delta t)^2$ with the constant $C$ independent of $\Delta x$, $\Delta t$, and $\delta$. Moreover, \eqref{eqn:linearmeta} yields
\begin{align*}
&\verti{\Delta x\sum_{\alpha=1}^d (\mathbf{B}\mathbf{C}_{test}-\mathbf{B}^\circ\mathbf{C}_{test}^\circ)_\alpha (\hat{u}(x_{i+\alpha},t^{n})+\hat{u}(x_{i-\alpha},t^{n})-2\hat{u}(x_{i},t^{n}))}\\
\leq&\Delta x\verti{\sum_{\alpha=1}^d (\mathbf{B}\mathbf{C}_{test}-\mathbf{B}^\circ\mathbf{C}_{test}^\circ)_\alpha(\alpha \Delta x)^2\max_{(x,t)\in\Omega_{pred}\times[0,T_{pred}]}\verti{\dfrac{\partial^2 \hat{u}}{\partial x^2}}}\\
\leq&\Delta x\delta^2\sum_{\alpha=1}^d \verti{(\mathbf{B}\mathbf{C}_{test}-\mathbf{B}^\circ\mathbf{C}_{test}^\circ)_\alpha}\max_{(x,t)\in\Omega_{pred}\times[0,T_{pred}]}\verti{\dfrac{\partial^2 \hat{u}}{\partial x^2}}\\
\leq&\Delta x\delta^2\sqrt{d}\vertii{\mathbf{B}\mathbf{C}_{test}-\mathbf{B}^\circ\mathbf{C}_{test}^\circ}_{l^2}\max_{(x,t)\in\Omega_{pred}\times[0,T_{pred}]}\verti{\dfrac{\partial^2 \hat{u}}{\partial x^2}}\\
\leq&\tilde{O}\left(\sqrt{\dfrac{\max\{H,d\} M^2}{N_{train}}+\dfrac{M}{N_{test}}}\right)\sqrt{\Delta x\delta^5}\max_{(x,t)\in\Omega_{pred}\times[0,T_{pred}]}\verti{\dfrac{\partial^2 \hat{u}}{\partial x^2}},
\end{align*}
with probability at least $1-O((\text{poly}(d))^{-1}+N_{test}^{-100})$. Hence we have the bound for $\epsilon_{all}$:
\begin{align*}
\verti{(\epsilon_{all})_i^n}\leq&E+\tilde{O}\left(\Delta x+(\Delta t)^2+\sqrt{\left(\dfrac{\max\{H,d\} M^2}{N_{train}}+\dfrac{M}{N_{test}}\right)\Delta x}\right).
\end{align*}

To show the $l^2(\omg)$ error for $e_i^n$, we first derive a bound for its error in the (discretized) energy norm. Multiplying \eqref{eqn:e1} with $\frac{e_i^{n+1}-e_i^{n}}{\Delta t}$ and summing over $\chi_{pred}=\{x_i\}_{i=1}^{L_{pred}}$ yields:
\begin{align*}
&\sum_{i=1}^{L_{pred}}\dfrac{(e_i^{n+1}-2e_i^{n}+e_i^{n-1})(e_i^{n+1}-e_i^{n})}{\Delta t^3}\\
= &\dfrac{\Delta x}{\Delta t}\sum_{i=1}^{L_{pred}}\sum_{\alpha=1}^d (\mathbf{B}^\circ\mathbf{C}_{test}^\circ)_\alpha (e_{i+\alpha}^{n}+e_{i-\alpha}^{n}-2e_{i}^{n})(e_i^{n+1}-e_i^{n})+\dfrac{1}{\Delta t}\sum_{i=1}^{L_{pred}}(\epsilon_{all})_i^n(e_i^{n+1}-e_i^{n}).   
\end{align*}
With the formulation $a(a-b)=\frac{1}{2}(a^2-b^2+(a-b)^2)$, we can rewrite the left hand side as
\begin{align*}
&\sum_{i=1}^{L_{pred}}\dfrac{(e_i^{n+1}-2e_i^{n}+e_i^{n-1})(e_i^{n+1}-e_i^{n})}{\Delta t^3}\\
\geq&\dfrac{1}{2\Delta t}\sum_{i=1}^{L_{pred}}\left[\left(\dfrac{e_i^{n+1}-e_i^{n}}{\Delta t}\right)^2-\left(\dfrac{e_i^{n}-e_i^{n-1}}{\Delta t}\right)^2+\left(\dfrac{e_i^{n+1}-2e_i^{n}+e_i^{n-1}}{\Delta t}\right)^2\right]\\
\geq&\dfrac{1}{2\Delta t}\sum_{i=1}^{L_{pred}}\left[\left(\dfrac{e_i^{n+1}-e_i^{n}}{\Delta t}\right)^2-\left(\dfrac{e_i^{n}-e_i^{n-1}}{\Delta t}\right)^2\right].
\end{align*}
For the first term on the right hand side, with the formulations
$$\sum_{i=1-\alpha}^{L} a_i(b_{i+\alpha}-b_i)=\sum_{i=1}^{\alpha}a_{L+i}b_{L+i}-\sum_{i=1}^{\alpha}a_{i-\alpha}b_{i-\alpha}-\sum_{i=1-\alpha}^{L}b_{i+\alpha}(a_{i+\alpha}-a_i),$$
$a(b-a)=\frac{1}{2}(b^2-a^2-(a-b)^2)$, Assumption \ref{asp:stab}, and the exact Dirichlet-type boundary condition, i.e., $e_i^n=0$ for $i<1$ and $i>L_{pred}$, we have
\begin{align*}
&\dfrac{\Delta x}{\Delta t}\sum_{i=1}^{L_{pred}}\sum_{\alpha=1}^d (\mathbf{B}^\circ\mathbf{C}_{test}^\circ)_\alpha (e_{i+\alpha}^{n}+e_{i-\alpha}^{n}-2e_{i}^{n})(e_i^{n+1}-e_i^{n})\\
=&-\dfrac{\Delta x}{\Delta t}\sum_{\alpha=1}^d\sum_{i=1-\alpha}^{L_{pred}} (\mathbf{B}^\circ\mathbf{C}_{test}^\circ)_\alpha (e^n_{i+\alpha}-e_i^n)(e^{n+1}_{i+\alpha}-e^n_{i+\alpha}-e_i^{n+1}+e_i^n)\\
=&-\dfrac{\Delta x}{2\Delta t}\sum_{\alpha=1}^d\sum_{i=1-\alpha}^{L_{pred}} (\mathbf{B}^\circ\mathbf{C}_{test}^\circ)_\alpha \left[(e^{n+1}_{i+\alpha}-e_i^{n+1})^2-(e^n_{i+\alpha}-e_i^n)^2\right.\\
&\left.-(e^{n+1}_{i+\alpha}-e^n_{i+\alpha}-e_i^{n+1}+e_i^n)^2\right]\\
\leq&-\dfrac{\Delta x}{2\Delta t}\sum_{\alpha=1}^d\sum_{i=1-\alpha}^{L_{pred}} (\mathbf{B}^\circ\mathbf{C}_{test}^\circ)_\alpha \left[(e^{n+1}_{i+\alpha}-e_i^{n+1})^2-(e^n_{i+\alpha}-e_i^n)^2\right]\\
&+\dfrac{\Delta x}{\Delta t}\sum_{\alpha=1}^d\sum_{i=1-\alpha}^{L_{pred}} (\mathbf{B}^\circ\mathbf{C}_{test}^\circ)_\alpha \left[(e^{n+1}_{i+\alpha}-e^n_{i+\alpha})^2+(e_i^{n+1}-e_i^n)^2\right]\\
\leq&-\dfrac{\Delta x}{2\Delta t}\sum_{\alpha=1}^d\sum_{i=1-\alpha}^{L_{pred}} (\mathbf{B}^\circ\mathbf{C}_{test}^\circ)_\alpha \left[(e^{n+1}_{i+\alpha}-e_i^{n+1})^2-(e^n_{i+\alpha}-e_i^n)^2\right]\\
&+2\dfrac{\Delta x}{\Delta t}\sum_{i=1}^{L_{pred}} \vertii{\mathbf{B}^\circ\mathbf{C}_{test}^\circ}_{l^1} (e_i^{n+1}-e_i^n)^2\\
\leq&-\dfrac{\Delta x}{2\Delta t}\sum_{\alpha=1}^d\sum_{i=1-\alpha}^{L_{pred}} (\mathbf{B}^\circ\mathbf{C}_{test}^\circ)_\alpha \left[(e^{n+1}_{i+\alpha}-e_i^{n+1})^2-(e^n_{i+\alpha}-e_i^n)^2\right]\\
&+\dfrac{1}{4}\sum_{i=1}^{L_{pred}} \left(\dfrac{e_i^{n+1}-e_i^n}{\Delta t}\right)^2.
\end{align*}
For the second term on the right hand side we have
\begin{align*}
\dfrac{1}{\Delta t}\sum_{i=1}^{L_{pred}}(\epsilon_{all})_i^n(e_i^{n+1}-e_i^{n})\leq&\sum_{i=1}^{L_{pred}}((\epsilon_{all})_i^n)^2+\dfrac{1}{4}\sum_{i=1}^{L_{pred}}\left(\dfrac{e_i^{n+1}-e_i^n}{\Delta t}\right)^2.    
\end{align*}
Putting the above three inequalities together, we obtain
\begin{align*}
&\sum_{i=1}^{L_{pred}}\left[\left(1-\Delta t\right)\left(\dfrac{e_i^{n+1}-e_i^{n}}{\Delta t}\right)^2-\left(\dfrac{e_i^{n}-e_i^{n-1}}{\Delta t}\right)^2\right]\\
&+2\Delta x\sum_{\alpha=1}^d\sum_{i=1-\alpha}^{L_{pred}} (\mathbf{B}^\circ\mathbf{C}_{test}^\circ)_\alpha \left[(e^{n+1}_{i+\alpha}-e_i^{n+1})^2-(e^n_{i+\alpha}-e_i^n)^2\right]\leq 2\Delta t\sum_{i=1}^{L_{pred}}((\epsilon_{all})_i^n)^2.
\end{align*}
With the discrete Gronwall lemma and the bound of $\Delta t$ in Assumption \ref{asp:stab}, for $n=1,\cdots,\lfloor T_{pred}/\Delta t\rfloor$ we have
\begin{align*}
&\sum_{i=1}^{L_{pred}}\left(\dfrac{e_i^{n}-e_i^{n-1}}{\Delta t}\right)^2+2\Delta x\sum_{\alpha=1}^d\sum_{i=1-\alpha}^{L_{pred}} (\mathbf{B}^\circ\mathbf{C}_{test}^\circ)_\alpha(e^{n}_{i+\alpha}-e_i^{n})^2\\
\leq &2L_{pred}((1-\Delta t)^{-n}-1)\max_{i,n}\verti{(\epsilon_{all})_i^n}^2\leq 4\exp(T_{pred})L_{pred}\max_{i,n}\verti{(\epsilon_{all})_i^n}^2\\
\leq& L_{pred}\tilde{O}\left(E^2+(\Delta x)^2+(\Delta t)^4+\left(\dfrac{\max\{H,d\} M^2}{N_{train}}+\dfrac{M}{N_{test}}\right)\Delta x\right),
\end{align*}
with probability at least $1-O((\text{poly}(d))^{-1}+N_{test}^{-100})$, which provides the error bound in the discrete energy norm. 
\end{proof}

\section{Reduction of Two Physics Constraints}

In this section we further expend the discussion on physics-based constraints in Section \ref{sec:physics}. The overall strategy is to fix the last two polynomial features, 
$$P_{M-1}(\xi)=\beta_1:=\left(\sum_{\alpha=1}^d\alpha^2\Delta x^3\right)^{-1}$$
and
$$P_{M}(\xi)=\beta_2\xi:=\left(\sum_{\alpha=1}^d\alpha^3\Delta x^4\right)^{-1}\xi$$
into the set of basis polynomials. We note that these two polynomials satisfy
$$\sum_{\alpha=1}^d \alpha^2 \Delta x^3 P_{M-1}(\alpha\Delta x)=1,\,\sum_{\alpha=1}^d \alpha^2 \Delta x^3 P_{M}(\alpha\Delta x)=1,$$
and
$$\sum_{\alpha=1}^d \alpha^4 \Delta x^5 P_{M-1}(\alpha\Delta x)=\dfrac{\sum_{\alpha=1}^d\alpha^4\Delta x^2}{\sum_{\alpha=1}^d\alpha^2},$$
$$\sum_{\alpha=1}^d \alpha^4 \Delta x^5 P_{M}(\alpha\Delta x)=\dfrac{\sum_{\alpha=1}^d\alpha^5\Delta x^2}{\sum_{\alpha=1}^d\alpha^3}.$$
Then \eqref{eqn:cond1_disc} writes
$$\bar{\rho} c_0^2=\sum_{m=1}^{M-2} C_m A_{1m}+C_{M-1}+C_{M},$$
and \eqref{eqn:cond2_disc} writes
$$-4\bar{\rho} c_0^3 R=\sum_{m=1}^{M-2} C_m A_{2m}+\dfrac{\sum_{\alpha=1}^d\alpha^4\Delta x^2}{\sum_{\alpha=1}^d\alpha^2}C_{M-1}+\dfrac{\sum_{\alpha=1}^d\alpha^5\Delta x^2}{\sum_{\alpha=1}^d\alpha^3}C_{M}.$$
Denoting 
\begin{equation*}
   \Lambda:= \begin{bmatrix}
   1 & 1 \\
\dfrac{\sum_{\alpha=1}^d\alpha^4\Delta x^2}{\sum_{\alpha=1}^d\alpha^2}&\dfrac{\sum_{\alpha=1}^d\alpha^5\Delta x^2}{\sum_{\alpha=1}^d\alpha^3}\\
\end{bmatrix},
\end{equation*}
and
\begin{equation*}
   \mathbf{H}:= \begin{bmatrix}
   \Delta x^2&4\Delta x^2&\cdots&d^2\Delta x^2\\
   \Delta x^4&16\Delta x^4&\cdots&d^4\Delta x^4\\
\end{bmatrix},
\end{equation*}
then
\begin{align*}
    \begin{bmatrix}
    C_{M-1} \\
    C_{M} \\
    \end{bmatrix}
    =& \Lambda^{-1}
    \begin{bmatrix}
    \bar{\rho} c_0^2 -  \sum_{m=1}^{M-2} C_m A_{1m}\\ 
    -4\bar{\rho} c_0^3 R-\sum_{m=1}^{M-2} C_m A_{2m} \\
    \end{bmatrix}\\
=&\Lambda^{-1}
    \begin{bmatrix}
    \bar{\rho} c_0^2\\ 
    -4\bar{\rho} c_0^3 R\\
    \end{bmatrix} -\Delta x\Lambda^{-1}\mathbf{H}\mathbf{B}\mathbf{C}.  
\end{align*}
Substituting this equation into the loss function in \eqref{eqn:linearr}, for each $x_i$ we obtain
\begin{align*}
&\left({y}_{k,i}^{n}-(\mathbf{s}_{k,i}^{n})^T\mathbf{B}\mathbf{C}\right)^2\\
=& \left({y}_{k,i}^{n}-(\mathbf{s}_{k,i}^{n})^T\left(\sum_{m=1}^{M-2} C_m\mathbf{b}_m +C_{M-1}\mathbf{b}_{M-1}+C_{M}\mathbf{b}_{M}\right)\right)^2\\
=&\left(y_{k,i}^{n}-(\mathbf{s}_{k,i}^{n})^T\sum_{m=1}^{M-2} C_m\mathbf{b}_m-(\mathbf{s}_{k,i}^{n})^T[\mathbf{b}_{M-1},\mathbf{b}_{M}]\Lambda^{-1}
\begin{bmatrix}
\bar{\rho} c_0^2\\
-4\bar{\rho} c_0^3 R\\
\end{bmatrix}\right.\\
&\left.+\Delta x(\mathbf{s}_{k,i}^{n})^T[\mathbf{b}_{M-1},\mathbf{b}_{M}]\Lambda^{-1}\mathbf{H}\mathbf{B}\mathbf{C}\right)^2\\
=&\left(y_{k,i}^{n}-(\mathbf{s}_{k,i}^{n})^T[\mathbf{b}_{M-1},\mathbf{b}_{M}]\Lambda^{-1}
\begin{bmatrix}
\bar{\rho} c_0^2\\
-4\bar{\rho} c_0^3 R\\
\end{bmatrix}\right.\\
&\left.-(\mathbf{s}_{k,i}^{n})^T (\mathbf{I}-\Delta x[\mathbf{b}_{M-1},\mathbf{b}_{M}]\Lambda^{-1}\mathbf{H}) \mathbf{B}\mathbf{C} \right)^2\\
=&\left(\tilde{y}_{k,i}^{n}-(\tilde{\mathbf{s}}_{k,i}^{n})^T{\mathbf{B}}\tilde{\mathbf{C}}\right)^2,
\end{align*}
where 
\begin{equation*}
\tilde{y}_{k,i}^{n}:=y_{k,i}^{n}-(\mathbf{s}_{k,i}^{n})^T[\mathbf{b}_{M-1},\mathbf{b}_{M}]\Lambda^{-1}
\begin{bmatrix}
\bar{\rho} c_0^2\\
-4\bar{\rho} c_0^3 R\\
\end{bmatrix},
\end{equation*}
$\tilde{\mathbf{s}}_{k,i}^{n}:=(\mathbf{I}-\Delta x[\mathbf{b}_{M-1},\mathbf{b}_{M}]\Lambda^{-1}\mathbf{H})^T\mathbf{s}_{k,i}^{n}$, $\tilde{\mathbf{C}}:=[C_1,\cdots,C_{M-2}]$, and $\mathbf{I}$ is an $d\times d$ identity matrix. Therefore, the analysis and algorithm can also be extended to the ``constraints'' cases.

\section{Detailed Parameter and Experiment Settings}

\subsection{Meta-train and Meta-test Datasets}

To demonstrate the performance of {\meta} on  both periodic and disordered materials, in empirical experiments we generate four types of data from the DNS solver for each microstructure. For each sample, the total training domain $\Omega=[-50,50]$ and the training data is generated up to $T = 2$. The spatial and
temporal discretization parameters in the DNS 
solver are set to $\Delta t = 0.01$, and $\max\verti{\Delta x} = 0.01$. The other physical parameters are set as $L=0.2$, $E_1=1$, $\rho_1=\rho_2=1$, and $\phi=0$. In experiment 1, we fix $E_2 = 0.25$ and set the disorder parameter $D\in [0.05,0.50]$. In experiment 2, we set $E_2 \in [0.2025,0.7225]$ and the disorder parameter $D=0$. The training and testing data are obtained from the DNS data via linear interpolation with $\Delta t = 0.02$ and $\Delta x = 0.05$. The two types of data are chosen to follow a similar setting as in \cite{You2021}:
\begin{enumerate}
    \item \textit{Oscillating source.} We let $\hat{u}(x,0) = \frac{\partial \hat{u}}{\partial t}(x,0) = 0$, $g(x,t) = \exp^{-(\frac{2x}{5kL})^2}\exp^{-(\frac{t-0.8}{0.8})^2}\cos^2(\frac{2\pi x}{kL})$, where $k=1,2,\cdots,20$. 
    \item \textit{Plane wave.}  We set $g(x,t) = 0$, $\hat{u}(x,-200) = 0$, and $\frac{\partial \hat{u}}{\partial t}(-50,t) = \cos(\omega t)$. In experiment 1 (random microstructures), we set $\omega  = 0.20, 0.40, \dots, 4.0$. In experiment 2 (periodic microstructures), we set $\omega = 0.30,0.60,\dots, 6.0$. 
\end{enumerate}
In these two types of loading scenarios, the displacement $\hat{u}(x,t)$ is mostly zero when $x>10$, which makes the corresponding datapoints carry very little information. To utilize the sample datapoints more efficiently, for the type 1 data we only use datapoints from the $x\in[-10,10]$ region, and for the type 2 data we only use datapoints from the $x\in[-38,-18]$ region.

\subsection{Validation Dataset: Wave Packet}

We create a validation dataset, denoted as the wave packet dataset, which considers a much longer bar ($\Omega_{wp}=[-133.3,133.3]$), and with a $50$ times longer simulation time ($t\in [0,100]$). The material is under a different loading condition from the training dataset, $g(x,t) = 0$ and $\frac{\partial \hat{u}}{\partial t}(-133.3, t) = \sin(\omega t) \exp\left(-(t/5 -3)^2\right)$, for $\omega =1,\, 2, \,3$. To provide a metric for the estimator accuracy, we calculate the averaged displacement error in the discretized energy norm at the last time step. This error metric is referred to as the ``validation error'', which checks the stability and generalizability of the estimators.

\subsection{Application: Projectile Impact Simulations}

To demonstrate the performance of learnt model in long term simulation, we simulate the long-term propagation of waves in this material due to the impact of a projectile. In particular, in this problem a projectile hits the left end of the bar at time zero, which generates a velocity wave that travels into the microstructures.

To demonstrate the generalization capability of our approach on different domains, boundary conditions, and longer simulation time, we consider a drastically different setting in this simulation task. In particular, a much larger domain, $\Omega_{impact} = (-267,267)$, and a much longer simulation time $t\in[0,600]$ are considered. Notice that our training dataset are only generated up to $T=2$, this long term simulation task is particularly challenging not only because it has a different boundary condition setting from all training samples, but also due to the large aspect ratio between training time scale and simulation time scale. On the left end of the domain, we prescribe the velocity as $\frac{\partial \hat{u}}{\partial t}(-267,0) = 1$, and zero velocity on elsewhere. 






\end{appendices}


\bibliography{references.bib}


\begin{thebibliography}{30}
\ifx \bisbn   \undefined \def \bisbn  #1{ISBN #1}\fi
\ifx \binits  \undefined \def \binits#1{#1}\fi
\ifx \bauthor  \undefined \def \bauthor#1{#1}\fi
\ifx \batitle  \undefined \def \batitle#1{#1}\fi
\ifx \bjtitle  \undefined \def \bjtitle#1{#1}\fi
\ifx \bvolume  \undefined \def \bvolume#1{\textbf{#1}}\fi
\ifx \byear  \undefined \def \byear#1{#1}\fi
\ifx \bissue  \undefined \def \bissue#1{#1}\fi
\ifx \bfpage  \undefined \def \bfpage#1{#1}\fi
\ifx \blpage  \undefined \def \blpage #1{#1}\fi
\ifx \burl  \undefined \def \burl#1{\textsf{#1}}\fi
\ifx \doiurl  \undefined \def \doiurl#1{\url{https://doi.org/#1}}\fi
\ifx \betal  \undefined \def \betal{\textit{et al.}}\fi
\ifx \binstitute  \undefined \def \binstitute#1{#1}\fi
\ifx \binstitutionaled  \undefined \def \binstitutionaled#1{#1}\fi
\ifx \bctitle  \undefined \def \bctitle#1{#1}\fi
\ifx \beditor  \undefined \def \beditor#1{#1}\fi
\ifx \bpublisher  \undefined \def \bpublisher#1{#1}\fi
\ifx \bbtitle  \undefined \def \bbtitle#1{#1}\fi
\ifx \bedition  \undefined \def \bedition#1{#1}\fi
\ifx \bseriesno  \undefined \def \bseriesno#1{#1}\fi
\ifx \blocation  \undefined \def \blocation#1{#1}\fi
\ifx \bsertitle  \undefined \def \bsertitle#1{#1}\fi
\ifx \bsnm \undefined \def \bsnm#1{#1}\fi
\ifx \bsuffix \undefined \def \bsuffix#1{#1}\fi
\ifx \bparticle \undefined \def \bparticle#1{#1}\fi
\ifx \barticle \undefined \def \barticle#1{#1}\fi
\bibcommenthead
\ifx \bconfdate \undefined \def \bconfdate #1{#1}\fi
\ifx \botherref \undefined \def \botherref #1{#1}\fi
\ifx \url \undefined \def \url#1{\textsf{#1}}\fi
\ifx \bchapter \undefined \def \bchapter#1{#1}\fi
\ifx \bbook \undefined \def \bbook#1{#1}\fi
\ifx \bcomment \undefined \def \bcomment#1{#1}\fi
\ifx \oauthor \undefined \def \oauthor#1{#1}\fi
\ifx \citeauthoryear \undefined \def \citeauthoryear#1{#1}\fi
\ifx \endbibitem  \undefined \def \endbibitem {}\fi
\ifx \bconflocation  \undefined \def \bconflocation#1{#1}\fi
\ifx \arxivurl  \undefined \def \arxivurl#1{\textsf{#1}}\fi
\csname PreBibitemsHook\endcsname

\bibitem{li2019design}
\begin{barticle}
\bauthor{\bsnm{Li}, \binits{Q.}},
\bauthor{\bsnm{He}, \binits{Z.}},
\bauthor{\bsnm{Li}, \binits{E.}},
\bauthor{\bsnm{Cheng}, \binits{A.}}:
\batitle{Design of a multi-resonator metamaterial for mitigating impact force}.
\bjtitle{Journal of Applied Physics}
\bvolume{125}(\bissue{3}),
\bfpage{035104}
(\byear{2019})
\end{barticle}
\endbibitem

\bibitem{yao2008experimental}
\begin{barticle}
\bauthor{\bsnm{Yao}, \binits{S.}},
\bauthor{\bsnm{Zhou}, \binits{X.}},
\bauthor{\bsnm{Hu}, \binits{G.}}:
\batitle{Experimental study on negative effective mass in a 1d mass--spring
  system}.
\bjtitle{New Journal of Physics}
\bvolume{10}(\bissue{4}),
\bfpage{043020}
(\byear{2008})
\end{barticle}
\endbibitem

\bibitem{huang2009wave}
\begin{barticle}
\bauthor{\bsnm{Huang}, \binits{H.}},
\bauthor{\bsnm{Sun}, \binits{C.}}:
\batitle{Wave attenuation mechanism in an acoustic metamaterial with negative
  effective mass density}.
\bjtitle{New Journal of Physics}
\bvolume{11}(\bissue{1}),
\bfpage{013003}
(\byear{2009})
\end{barticle}
\endbibitem

\bibitem{manimala2014dynamic}
\begin{barticle}
\bauthor{\bsnm{Manimala}, \binits{J.M.}},
\bauthor{\bsnm{Huang}, \binits{H.H.}},
\bauthor{\bsnm{Sun}, \binits{C.}},
\bauthor{\bsnm{Snyder}, \binits{R.}},
\bauthor{\bsnm{Bland}, \binits{S.}}:
\batitle{Dynamic load mitigation using negative effective mass structures}.
\bjtitle{Engineering structures}
\bvolume{80},
\bfpage{458}--\blpage{468}
(\byear{2014})
\end{barticle}
\endbibitem

\bibitem{du2019multiscale}
\begin{botherref}
\oauthor{\bsnm{Du}, \binits{Q.}},
\oauthor{\bsnm{Engquist}, \binits{B.}},
\oauthor{\bsnm{Tian}, \binits{X.}}:
Multiscale modeling, homogenization and nonlocal effects: Mathematical and
  computational issues
(2019)
\end{botherref}
\endbibitem

\bibitem{barker1971model}
\begin{barticle}
\bauthor{\bsnm{Barker}, \binits{L.}}:
\batitle{A model for stress wave propagation in composite materials}.
\bjtitle{Journal of Composite Materials}
\bvolume{5}(\bissue{2}),
\bfpage{140}--\blpage{162}
(\byear{1971})
\end{barticle}
\endbibitem

\bibitem{You2021}
\begin{botherref}
\oauthor{\bsnm{You}, \binits{H.}},
\oauthor{\bsnm{Yu}, \binits{Y.}},
\oauthor{\bsnm{Silling}, \binits{S.}},
\oauthor{\bsnm{D'Elia}, \binits{M.}}:
Data-driven learning of nonlocal models: from high-fidelity simulations to
  constitutive laws.
AAAI Spring Symposium: MLPS
(2021)
\end{botherref}
\endbibitem

\bibitem{silling2021propagation}
\begin{botherref}
\oauthor{\bsnm{Silling}, \binits{S.A.}}:
Propagation of a stress pulse in a heterogeneous elastic bar.
Journal of Peridynamics and Nonlocal Modeling,
1--21
(2021)
\end{botherref}
\endbibitem

\bibitem{deshmukh2021}
\begin{botherref}
\oauthor{\bsnm{Deshmukh}, \binits{K.}},
\oauthor{\bsnm{Breitzman}, \binits{T.}},
\oauthor{\bsnm{Dayal}, \binits{K.}}:
Multiband homogenization of metamaterials in real-space: Higher-order nonlocal
  models and scattering at external surface.
preprint
(2021)
\end{botherref}
\endbibitem

\bibitem{You2020Regression}
\begin{barticle}
\bauthor{\bsnm{You}, \binits{H.}},
\bauthor{\bsnm{Yu}, \binits{Y.}},
\bauthor{\bsnm{Trask}, \binits{N.}},
\bauthor{\bsnm{Gulian}, \binits{M.}},
\bauthor{\bsnm{D'Elia}, \binits{M.}}:
\batitle{Data-driven learning of robust nonlocal physics from high-fidelity
  synthetic data}.
\bjtitle{Computer Methods in Applied Mechnics and Engineering}
\bvolume{374},
\bfpage{113553}
(\byear{2021})
\end{barticle}
\endbibitem

\bibitem{you2021md}
\begin{barticle}
\bauthor{\bsnm{You}, \binits{H.}},
\bauthor{\bsnm{Yu}, \binits{Y.}},
\bauthor{\bsnm{Silling}, \binits{S.}},
\bauthor{\bsnm{D’Elia}, \binits{M.}}:
\batitle{A data-driven peridynamic continuum model for upscaling molecular
  dynamics}.
\bjtitle{Computer Methods in Applied Mechanics and Engineering}
\bvolume{389},
\bfpage{114400}
(\byear{2022})
\end{barticle}
\endbibitem

\bibitem{tripuraneni2021provable}
\begin{bchapter}
\bauthor{\bsnm{Tripuraneni}, \binits{N.}},
\bauthor{\bsnm{Jin}, \binits{C.}},
\bauthor{\bsnm{Jordan}, \binits{M.}}:
\bctitle{Provable meta-learning of linear representations}.
In: \bbtitle{International Conference on Machine Learning},
pp. \bfpage{10434}--\blpage{10443}
(\byear{2021}).
\bcomment{PMLR}
\end{bchapter}
\endbibitem

\bibitem{beran1970mean}
\begin{barticle}
\bauthor{\bsnm{Beran}, \binits{M.}},
\bauthor{\bsnm{McCoy}, \binits{J.}}:
\batitle{Mean field variations in a statistical sample of heterogeneous
  linearly elastic solids}.
\bjtitle{International Journal of Solids and Structures}
\bvolume{6}(\bissue{8}),
\bfpage{1035}--\blpage{1054}
(\byear{1970})
\end{barticle}
\endbibitem

\bibitem{silling2014origin}
\begin{barticle}
\bauthor{\bsnm{Silling}, \binits{S.A.}}:
\batitle{Origin and effect of nonlocality in a composite}.
\bjtitle{Journal of Mechanics of Materials and Structures}
\bvolume{9}(\bissue{2}),
\bfpage{245}--\blpage{258}
(\byear{2014})
\end{barticle}
\endbibitem

\bibitem{du2020multiscale}
\begin{botherref}
\oauthor{\bsnm{Du}, \binits{Q.}},
\oauthor{\bsnm{Engquist}, \binits{B.}},
\oauthor{\bsnm{Tian}, \binits{X.}}:
Multiscale modeling, homogenization and nonlocal effects: Mathematical and
  computational issues.
Contemporary mathematics
\textbf{754}
(2020)
\end{botherref}
\endbibitem

\bibitem{du2017peridynamic}
\begin{botherref}
\oauthor{\bsnm{Du}, \binits{Q.}},
\oauthor{\bsnm{Tao}, \binits{Y.}},
\oauthor{\bsnm{Tian}, \binits{X.}}:
A peridynamic model of fracture mechanics with bond-breaking.
Journal of Elasticity,
1--22
(2017)
\end{botherref}
\endbibitem

\bibitem{ge2017no}
\begin{bchapter}
\bauthor{\bsnm{Ge}, \binits{R.}},
\bauthor{\bsnm{Jin}, \binits{C.}},
\bauthor{\bsnm{Zheng}, \binits{Y.}}:
\bctitle{No spurious local minima in nonconvex low rank problems: A unified
  geometric analysis}.
In: \bbtitle{International Conference on Machine Learning},
pp. \bfpage{1233}--\blpage{1242}
(\byear{2017}).
\bcomment{PMLR}
\end{bchapter}
\endbibitem

\bibitem{liu2017materials}
\begin{barticle}
\bauthor{\bsnm{Liu}, \binits{Y.}},
\bauthor{\bsnm{Zhao}, \binits{T.}},
\bauthor{\bsnm{Ju}, \binits{W.}},
\bauthor{\bsnm{Shi}, \binits{S.}}:
\batitle{Materials discovery and design using machine learning}.
\bjtitle{Journal of Materiomics}
\bvolume{3}(\bissue{3}),
\bfpage{159}--\blpage{177}
(\byear{2017})
\end{barticle}
\endbibitem

\bibitem{lu2018accelerated}
\begin{barticle}
\bauthor{\bsnm{Lu}, \binits{S.}},
\bauthor{\bsnm{Zhou}, \binits{Q.}},
\bauthor{\bsnm{Ouyang}, \binits{Y.}},
\bauthor{\bsnm{Guo}, \binits{Y.}},
\bauthor{\bsnm{Li}, \binits{Q.}},
\bauthor{\bsnm{Wang}, \binits{J.}}:
\batitle{Accelerated discovery of stable lead-free hybrid organic-inorganic
  perovskites via machine learning}.
\bjtitle{Nature communications}
\bvolume{9}(\bissue{1}),
\bfpage{1}--\blpage{8}
(\byear{2018})
\end{barticle}
\endbibitem

\bibitem{butler2018machine}
\begin{barticle}
\bauthor{\bsnm{Butler}, \binits{K.T.}},
\bauthor{\bsnm{Davies}, \binits{D.W.}},
\bauthor{\bsnm{Cartwright}, \binits{H.}},
\bauthor{\bsnm{Isayev}, \binits{O.}},
\bauthor{\bsnm{Walsh}, \binits{A.}}:
\batitle{Machine learning for molecular and materials science}.
\bjtitle{Nature}
\bvolume{559}(\bissue{7715}),
\bfpage{547}--\blpage{555}
(\byear{2018})
\end{barticle}
\endbibitem

\bibitem{cai2020machine}
\begin{barticle}
\bauthor{\bsnm{Cai}, \binits{J.}},
\bauthor{\bsnm{Chu}, \binits{X.}},
\bauthor{\bsnm{Xu}, \binits{K.}},
\bauthor{\bsnm{Li}, \binits{H.}},
\bauthor{\bsnm{Wei}, \binits{J.}}:
\batitle{Machine learning-driven new material discovery}.
\bjtitle{Nanoscale Advances}
\bvolume{2}(\bissue{8}),
\bfpage{3115}--\blpage{3130}
(\byear{2020})
\end{barticle}
\endbibitem

\bibitem{iwasaki2019machine}
\begin{barticle}
\bauthor{\bsnm{Iwasaki}, \binits{Y.}},
\bauthor{\bsnm{Takeuchi}, \binits{I.}},
\bauthor{\bsnm{Stanev}, \binits{V.}},
\bauthor{\bsnm{Kusne}, \binits{A.G.}},
\bauthor{\bsnm{Ishida}, \binits{M.}},
\bauthor{\bsnm{Kirihara}, \binits{A.}},
\bauthor{\bsnm{Ihara}, \binits{K.}},
\bauthor{\bsnm{Sawada}, \binits{R.}},
\bauthor{\bsnm{Terashima}, \binits{K.}},
\bauthor{\bsnm{Someya}, \binits{H.}}, \betal:
\batitle{Machine-learning guided discovery of a new thermoelectric material}.
\bjtitle{Scientific reports}
\bvolume{9}(\bissue{1}),
\bfpage{1}--\blpage{7}
(\byear{2019})
\end{barticle}
\endbibitem

\bibitem{ren2018accelerated}
\begin{barticle}
\bauthor{\bsnm{Ren}, \binits{F.}},
\bauthor{\bsnm{Ward}, \binits{L.}},
\bauthor{\bsnm{Williams}, \binits{T.}},
\bauthor{\bsnm{Laws}, \binits{K.J.}},
\bauthor{\bsnm{Wolverton}, \binits{C.}},
\bauthor{\bsnm{Hattrick-Simpers}, \binits{J.}},
\bauthor{\bsnm{Mehta}, \binits{A.}}:
\batitle{Accelerated discovery of metallic glasses through iteration of machine
  learning and high-throughput experiments}.
\bjtitle{Science advances}
\bvolume{4}(\bissue{4}),
\bfpage{1566}
(\byear{2018})
\end{barticle}
\endbibitem

\bibitem{kaufmann2020discovery}
\begin{barticle}
\bauthor{\bsnm{Kaufmann}, \binits{K.}},
\bauthor{\bsnm{Maryanovsky}, \binits{D.}},
\bauthor{\bsnm{Mellor}, \binits{W.M.}},
\bauthor{\bsnm{Zhu}, \binits{C.}},
\bauthor{\bsnm{Rosengarten}, \binits{A.S.}},
\bauthor{\bsnm{Harrington}, \binits{T.J.}},
\bauthor{\bsnm{Oses}, \binits{C.}},
\bauthor{\bsnm{Toher}, \binits{C.}},
\bauthor{\bsnm{Curtarolo}, \binits{S.}},
\bauthor{\bsnm{Vecchio}, \binits{K.S.}}:
\batitle{Discovery of high-entropy ceramics via machine learning}.
\bjtitle{Npj Computational Materials}
\bvolume{6}(\bissue{1}),
\bfpage{1}--\blpage{9}
(\byear{2020})
\end{barticle}
\endbibitem

\bibitem{finn2017model}
\begin{bchapter}
\bauthor{\bsnm{Finn}, \binits{C.}},
\bauthor{\bsnm{Abbeel}, \binits{P.}},
\bauthor{\bsnm{Levine}, \binits{S.}}:
\bctitle{Model-agnostic meta-learning for fast adaptation of deep networks}.
In: \bbtitle{International Conference on Machine Learning},
pp. \bfpage{1126}--\blpage{1135}
(\byear{2017}).
\bcomment{PMLR}
\end{bchapter}
\endbibitem

\bibitem{qiu2020meta}
\begin{barticle}
\bauthor{\bsnm{Qiu}, \binits{Y.L.}},
\bauthor{\bsnm{Zheng}, \binits{H.}},
\bauthor{\bsnm{Devos}, \binits{A.}},
\bauthor{\bsnm{Selby}, \binits{H.}},
\bauthor{\bsnm{Gevaert}, \binits{O.}}:
\batitle{A meta-learning approach for genomic survival analysis}.
\bjtitle{Nature communications}
\bvolume{11}(\bissue{1}),
\bfpage{1}--\blpage{11}
(\byear{2020})
\end{barticle}
\endbibitem

\bibitem{chen2021metadelta}
\begin{botherref}
\oauthor{\bsnm{Chen}, \binits{Y.}},
\oauthor{\bsnm{Guan}, \binits{C.}},
\oauthor{\bsnm{Wei}, \binits{Z.}},
\oauthor{\bsnm{Wang}, \binits{X.}},
\oauthor{\bsnm{Zhu}, \binits{W.}}:
Metadelta: A meta-learning system for few-shot image classification.
arXiv preprint arXiv:2102.10744
(2021)
\end{botherref}
\endbibitem

\bibitem{yin2020meta}
\begin{botherref}
\oauthor{\bsnm{Yin}, \binits{W.}}:
Meta-learning for few-shot natural language processing: A survey.
arXiv preprint arXiv:2007.09604
(2020)
\end{botherref}
\endbibitem

\bibitem{kailkhura2019reliable}
\begin{barticle}
\bauthor{\bsnm{Kailkhura}, \binits{B.}},
\bauthor{\bsnm{Gallagher}, \binits{B.}},
\bauthor{\bsnm{Kim}, \binits{S.}},
\bauthor{\bsnm{Hiszpanski}, \binits{A.}},
\bauthor{\bsnm{Han}, \binits{T.Y.-J.}}:
\batitle{Reliable and explainable machine-learning methods for accelerated
  material discovery}.
\bjtitle{npj Computational Materials}
\bvolume{5}(\bissue{1}),
\bfpage{1}--\blpage{9}
(\byear{2019})
\end{barticle}
\endbibitem

\bibitem{mai2021use}
\begin{botherref}
\oauthor{\bsnm{Mai}, \binits{H.}},
\oauthor{\bsnm{Le}, \binits{T.C.}},
\oauthor{\bsnm{Hisatomi}, \binits{T.}},
\oauthor{\bsnm{Chen}, \binits{D.}},
\oauthor{\bsnm{Domen}, \binits{K.}},
\oauthor{\bsnm{Winkler}, \binits{D.A.}},
\oauthor{\bsnm{Caruso}, \binits{R.A.}}:
Use of meta models for rapid discovery of narrow bandgap oxide photocatalysts.
iScience,
103068
(2021)
\end{botherref}
\endbibitem

\end{thebibliography}


\end{document}